\documentclass[
 reprint,
 amsmath,amssymb,
 aps,
 floatfix,
 prx,
 superscriptaddress
]{revtex4-2}

\usepackage{graphicx}
\usepackage{dcolumn}
\usepackage{bm}
\usepackage{comment}
\usepackage{color}
\usepackage{float}
\usepackage[caption=false]{subfig}

\newcommand{\R}{\mathbb{R}}

\newcommand{\Z}{\mathbb{Z}}

\newcommand{\pd}[2]{\frac{\partial #1}{\partial #2}}

\newcommand*{\dd}{\mathrm{d}}

\usepackage[margin=30truemm]{geometry}

\begin{document}

\preprint{APS/123-QED}

\title{Reconstructing the information processing capacity of physical systems from noisy observations}

\author{Shun Kotoku}
\affiliation{%
 Department of Information Physics and Computing, Graduate School of Information Science and Technology,
The University of Tokyo, 7-3-1 Hongo, Bunkyo-ku, Tokyo 113-8656, Japan
}%
\author{Rodrigo Mart\'{\i}nez-Pe\~{n}a}
\affiliation{
 Donostia International Physics Center, Paseo Manuel de Lardizabal 4, E-20018 San Sebasti\'{a}n, Spain
}
\author{Takatomo Mihana}
\affiliation{%
 Department of Information Physics and Computing, Graduate School of Information Science and Technology,
The University of Tokyo, 7-3-1 Hongo, Bunkyo-ku, Tokyo 113-8656, Japan
}%
\author{Felix Köster}
\affiliation{%
 Department of Information and Computer Sciences, Saitama University, 255 Shimo-Okubo, Sakura-ku, Saitama 338-8570, Japan
}%
\author{Johannes Nokkala}
\affiliation{
 Department of Physics and Astronomy, University of Turku, FI-20014 Turun yliopisto, Finland
}
\affiliation{
 Turku Collegium for Science, Medicine and Technology, University of Turku, Turku, Finland
}
\author{Lina Jaurigue}
\affiliation{
 Institute of Physics, Technische Universitat Ilmenau, P.O.Box 100565, D-98684 Ilmenau, Germany
}
\author{Ryoichi Horisaki}
\affiliation{%
 Department of Information Physics and Computing, Graduate School of Information Science and Technology,
The University of Tokyo, 7-3-1 Hongo, Bunkyo-ku, Tokyo 113-8656, Japan
}%
\author{Andr\'{e} R\"{o}hm}%
\email{roehm@g.ecc.u-tokyo.ac.jp}
\affiliation{%
 Department of Information Physics and Computing, Graduate School of Information Science and Technology,
The University of Tokyo, 7-3-1 Hongo, Bunkyo-ku, Tokyo 113-8656, Japan
}%

\date{\today}

\begin{abstract}

Driven dynamical systems can compute when their transient states encode complex transformations of past inputs.
The information processing capacity (IPC) framework allows for a detailed accounting of these computational properties, however its interpretation in noisy systems has remained incomplete. 
In this work, we clarify how noise affects the IPC and how one can reconstruct the noiseless IPC.
First, we show how to distinguish the dynamics of an unperturbed system from the noise-free component of the stochastic dynamics: The IPC measured for responses averaged over noise realizations is in general not the same as the IPC of the unperturbed system. 
We explicitly demonstrate that noise can redistribute computational capacity and sometimes even enhance performance on particular tasks, rather than merely degrading a fixed computation.
We then introduce covariance reconstruction by orthogonal projection (CROP), which reconstructs the covariance and IPC of the noise-free component directly from noisy observations, without requiring a detailed model of either the system or the noise.
At fixed total measurement budget, numerical tests on a classical nonlinear reservoir show that CROP estimates the noise-free IPC more accurately than the standard practice of ensemble averaging over repeated trials. 
We find the same advantage in a quantum reservoir subject to unavoidable measurement noise. 
Our results provide a general route to recovering the computational structure of noisy physical systems from finite observations.
\end{abstract}

\maketitle


\section{Introduction}
Recent years have seen a surge of interest in computing approaches that go beyond the classic silicon-transistor based digital models.
These approaches aim to utilize the computational power of physical processes without necessarily relying on boolean logic or gates.
This includes physical neural networks \cite{WrightOnodera2022,momeni2023backpropagation,yu2024physical,sunada2025blending}, photonic computing \cite{McMahon2023, Shastri2021, Abreu2024}, reservoir computing \cite{gallicchio2017deep,van2017advances, Tanaka2019,nakajima2021reservoir} and  spintronic platforms \cite{torrejon2017neuromorphic,Grollier2020,chen2023spintronic}.
Both the physics of classical phenomena as well as quantum physics can be used, the latter of which is now at the center of the quickly growing field of quantum reservoir computing \cite{mujal2021opportunities,Neuromorphic2021rev,fujii2021quantum,abbas2024classical,CIN24}.

Unlike the explicit algorithm-design of digital computers, a general theory of computing with arbitrary physical systems is still lacking \cite{Jaeger2023, Crutchfield2012}.
Therefore, designing such systems often involves optimizing network structures or a large number of parameters.
This process often involves trial-and-error or iterative routines such as backpropagation. 
Consequently, insights obtained from one system are rarely transferable to different architectures or tasks, underscoring the lack of systematic design principles. 

The information processing capacity (IPC)~\cite{Dambre2012} is an established framework for quantifying the computational power of input-driven dynamical systems. 
As detailed in the following sections, the IPC serves as a natural metric by enabling a task-independent evaluation of how the system retains and transforms past inputs into specific basis functions. 
This framework is essentially equivalent to the Wiener-Volterra series expansion~\cite{Kubota2021}, providing a systematic and transparent view of the system's capabilities. 
Thereby, to a certain extent \cite{HUE22a}, addressing the inherent limitations of black-box optimization.

When utilizing real-world physical substrates, systems are inevitably subject to noise that interferes with their computational processes. 
While certain noise levels can occasionally enrich system dynamics, e.g., through stochastic resonance~\cite{Wiesenfeld1995, Hanggi2002}, noise is, in general, understood as an inaccessible signal component that degrades the fidelity of information processing. 
The degradation of the IPC due to noise has been investigated in previous studies~\cite{Dambre2012, Vettelschoss2021}, and was recently formally characterized~\cite{Polloreno2023}. 
This, importantly, also covers the special case of the measurement process in quantum mechanics, where an ensemble of copies of the system is required to compute its observables \cite{hu2023tackling,vcindrak2026memory}.

However, the existing theory only allows the estimation of how much the IPC of a dynamical system is reduced by noise, starting from the unperturbed system. 
In practice, experimental physical systems always contain noise. 
Thus, a commonly adopted practice is to rerun the same "computation" multiple times, and average state traces to increase the signal-to-noise ratio. 
This is especially common for quantum reservoir computing \cite{mujal2023time,yasuda2023quantum,hu2024overcoming}. 
Predicting from the original noisy data what the IPC will become after noise-reduction strategies (such as averaging) have been applied, is an open problem.

In this paper, we clarify and relax the assumptions regarding noise-induced deviations, thereby extending the previous formulation of noise effects on the IPC and enabling its application to a wider range of noisy dynamical systems. 
First, we clarify that noise not only reduces the total IPC, but can also shift the dynamics of the system. 
Here, we carefully differentiate between the unperturbed dynamics without noise, and the noise-free part of the stochastic system.
While in most numerically tested cases, these two IPC distributions are close, they are in general not identical.
In particular, we show an example where noise increases the performance on a particular task. 

We then propose a method to estimate the noise-free part of the IPC purely from noisy observations. 
We call this method Covariance Reconstruction by Orthogonal Projection (CROP).
Such an estimation is useful, as it allows for an assessment of the IPC distribution if the noise was averaged out.
Consequently, for physical platforms where noise is fundamentally unavoidable, such as in the context of quantum reservoir computing, the estimated noise-free IPC serves as a systematic benchmark to efficiently identify promising physical substrates, network architectures, or parameter regimes. 
Notably, this systematic decomposition of computational capacity offers insights into the system's computational capability that conventional task-dependent evaluations cannot provide. 
We then show that CROP is more sample-efficient for estimating the IPC than current standard noise-reduction methods.

The remainder of this paper is organized as follows. In Sec.~\ref{sec:theory}, we introduce the IPC framework for stochastic dynamical systems, distinguish the unperturbed dynamics from the noise-free component of a noisy system, and derive the CROP method in Sec.~\ref{subsec:proposed} for estimating the corresponding noise-free IPC. In Sec.~\ref{sec:classical}, we apply CROP to a classical reservoir with quadratic nonlinearity and compare its estimates with analytical results and conventional ensemble averaging. In Sec.~\ref{sec:quantum}, we demonstrate the method for a quantum reservoir subject to measurement-induced shot noise and examine the trade-off between the number of measurement shots and the available time-series length. Finally, Sec.~\ref{sec:conclusion} summarizes our results and discusses their implications.

\section{Theory}
\label{sec:theory}


\subsection{Representation of dynamical systems}
\label{subsec:represent}

We consider dynamical systems driven by an external input. 
In the discrete-time formulation, the dynamical state evolves according to the map 
\begin{align}
    r[k] = f(r[k-1], u[k]), \quad k\in\Z,
    \label{eq:map}
\end{align}
where $r[k] := (r_1[k], \ldots, r_N[k])^{\mathsf{T}}\in\R^{N}$ denotes the system state at time step $k$ and $u[k] := (u_1[k], \ldots, u_M[k])^{\mathsf{T}}\in\mathcal{U}\subseteq\R^M$ denotes the input signal, where $\mathcal{U}$ is the support of the input distribution. 
Here $N$ and $M$ represent the dimensions of the dynamical state and the input space, respectively, whereas 
$f: \R^{N}\times \R^M\to\R^{N}$ is the state update function. 
Other input-driven dynamical systems can be described in continuous time by a differential equation of the form 
\begin{align}
    \frac{\dd r(t)}{\dd t} = g(r(t), u(t)), \quad t\in\R,
    \label{eq:de}
\end{align}
where $r(t)\in\R^{N}$ and $u(t)\in\R^M$ denote the time-continuous state and input at time $t$, respectively. 
$g: \R^{N}\times\R^M\to\R^N$ represents a (possibly nonlinear) vector field. 
In practice, continuous-time dynamics are not accessed directly but are observed at a finite sampling time $T_{\mathrm{samp}}$. 
The discretely observed state is therefore given by 
\begin{align}
    r[k] = r(kT_{\mathrm{samp}}). 
\end{align}
In this context, the discrete-time map $f$ in Eq.~\eqref{eq:map} can be viewed as the flow map of Eq.~\eqref{eq:de}, which maps the state $r((k-1)T_{\mathrm{samp}})$ to $r(kT_{\mathrm{samp}})$ by integrating the vector field $g$ over the time interval $[(k-1)T_{\mathrm{samp}}, kT_{\mathrm{samp}})$. 
Moreover, when the input-driving system is assumed to have a finite bandwidth, we consider a piecewise-constant input of the form 
\begin{align}
    u(t) = u[k], \quad (k-1) \, T_{\mathrm{samp}} \le t < k \,T_{\mathrm{samp}}.
\end{align}

While we mainly use the discrete-time formulation in Eq.~\eqref{eq:map} for simplicity, the methods presented in this paper also apply to other dynamical systems, including ordinary differential equations, delay differential equations, and iterative maps or hysteresis operators \cite{JAU25a}. 

The formulations presented so far describe systems driven solely by the input signal $u$. 
In practice, however, physical substrates are inevitably subject to stochastic perturbations. 
In this work we explore how noise degrades the performance and information processing in input-driven systems.

Noise can affect different parts of the system. 
It can occur on the input, inside the state updates, or at the final measurements.
To formalize these scenarios, let $r'[k]\in\R^N$ denote the observed system state under the influence of noise. 
First, in the case of additive process noise $v[k]\in\R^{N}$, the dynamics are described by
\begin{align}
    r'[k] = f\left(r'[k-1], u[k]\right) + v[k].
    \label{eq:resnoise}
\end{align}
More complex forms of process noise are also possible, including multiplicative noise or other modifications of $f$ as
\begin{align}
    r'[k] = f'\left(r'[k-1], u[k], v[k]\right),
    \label{eq:resnoise_general}
\end{align}
where $f'\left(r'[k-1], u[k], 0\right) = f\left(r'[k-1], u[k]\right)$.
Secondly, the noise can occur at the point of entry into the dynamical system, leading to dynamics of the form 
\begin{align}
    r'[k] = f\left(r'[k-1], u[k] + v^{\mathrm{in}}[k]\right),
    \label{eq:innoise}
\end{align}
where $v^{\mathrm{in}}[k]\in\R^M$ denotes the input noise. 
Finally, for output noise, the system state is affected at the read-out stage. 
In this case, the observed state is given by
\begin{align}
    r'[k] = r[k] + v^{\mathrm{out}}[k],
    \label{eq:outnoise}
\end{align}
where $v^{\mathrm{out}}[k]\in\R^{N}$ represents additive noise on the observed state. 

Together, these models capture a broad range of stochastic perturbations in dynamical systems, including thermal fluctuations in physical components and shot noise or sensing errors in read-out interfaces. 

\subsection{Mutual information}
To characterize the input-driven stochastic dynamical systems formulated in Sec.~\ref{subsec:represent}, one might turn to information-theoretic metrics. 
A conventional approach involves calculating the mutual information (MI) between the $h$-step input history $u^{(h)}[k] := (u[k], u[k-1], \ldots u[k-h+1])\in\mathcal{U}^{h}$ and the system state $r[k]\in\R^{N}$. 
MI is defined as 
\begin{align}
    &I(u^{(h)}[k]; r[k]) \notag := H(u^{(h)}[k]) - H(u^{(h)}[k]\mid r[k]), 
\end{align}
where $H(\cdot)$ denotes the differential entropy. 
Intuitively, mutual information quantifies how much information the system state $r[k]$ and the input history $u^{(h)}[k]$ share, or equivalently, how much uncertainty about one is reduced by observing the other. 
Indeed, when one applies such a framework to systems with noise, as the strength of the noise $v[k]$ increases, the MI between input and system state will decrease, indicating a loss of information about the inputs.

However, MI quantifies only the total amount of statistical dependence and, by itself, does not reveal how this information is represented in terms of memory or nonlinearity.

For example, consider the following two systems driven by a one-dimensional input $(M=1)$ under noise corrupted input: 
A scalar ($N=1$) system that stores a linear copy of the last input
\begin{align}
    r_{\mathrm{A}}'[k] &= \begin{pmatrix}
       u[k] +  v[k]
    \end{pmatrix},\notag\\
    \label{eq:rA}
\end{align}
and a larger $(N \geq 2$) system that stores higher powers of the same (where the noise per row is not independent):
\begin{align}
    r_{\mathrm{B}}'[k] = \begin{pmatrix}
       (u[k] + v[k])\\
       (u[k] + v[k])^2 \\
       \dots  \\
       (u[k] + v[k])^N
    \end{pmatrix}.\notag\\
    \label{eq:rB}
\end{align}
Both systems yield identical MI, $I(u^{(h)}[k]; r_{\mathrm{A}}'[k]) = I(u^{(h)}[k]; r_{\mathrm{B}}'[k])$, because the MI is invariant under invertible transformations. Every additional row of $r_{\mathrm{B}}'[k]$ can in principle be deduced from the first row. 
However, intuitively, systems A and B represent qualitatively different information about the input history. System A stores a single distinct representation, whereas System B clearly implements a wider variety of computation. 
This reveals a key limitation of MI; it measures only the total amount of information, while ignoring how that information is encoded. 
Redundant information is not directly accounted for, even if highly transformed.
Therefore, a finer-grained description is required to characterize the specific computations realized by a dynamical system.

\subsection{Information processing capacity (IPC)\label{subsec:ipc}}
The MI quantifies how well we can reconstruct the input $u[k]$ from the system state $r[k]$. 
However, MI allows nonlinear transformations for this reconstruction, which are potentially more computationally complex than the system itself.
Thus, we want to more carefully separate different representations of information.
For this, we will follow the well-known IPC framework as introduced by Dambre \textit{et.~al} \cite{Dambre2012}.

We quantify a driven dynamical system's computational capability by evaluating how accurately affine projections of its state can reconstruct a set of target functions $\{z\}$.
These functions, defined in terms of the input history $u^{(h)}[k]$, serve as a basis for representing the information processing performed by the system. 
To isolate the system's intrinsic computation from correlations in the input, we assume that the input sequence $\{u[k]\}_{k\in\Z}$ is independent and identically distributed (i.i.d.). 

We consider arbitrary target functions $z(u^{(h)}[k])$, where $z: \mathcal{U}^h\to\R$. 
Let $\mu_u$ denote the probability measure induced by the input history $u^{(h)}[k]$, and define the expectation operator (for measurable $z$)
\begin{align}
    \langle z\rangle_u := \int_{\mathcal{U}^h}z(\omega)d\mu_u(\omega), 
\end{align}
where $\omega\in\mathcal{U}^h$ denotes a realization of the input history. 
We introduce the Hilbert space of input-output functions
\begin{align}
    \mathcal{H}_u := \left\{z: \mathcal{U}^h\to\R\mid \langle z^2\rangle_u < \infty\right\},
\end{align}
equipped with the inner product
\begin{align}
    \langle z,\zeta\rangle_u := \langle z\zeta\rangle_u,
\end{align}
and the induced norm
\begin{align}
    \|z\|_u := \sqrt{\langle z, z\rangle_u} = \sqrt{\langle z^2\rangle_u}. 
\end{align}

We assume that the dynamical system $f(\dots)$ in Eq.~\eqref{eq:map} satisfies the echo state property, so that the system state is uniquely determined by the entire past input sequence. 
For clarity, we restrict our attention to a sufficiently long finite history $u^{(h)}[k]$ --- although formally, the IPC theory assumes the limit of $h \rightarrow \infty$ --- and regard the system state as being determined by this truncated history
\begin{align}
    r[k] = R(u^{(h)}[k]), \quad R: \mathcal{U}^h\to\R^N,
    \label{eq:reservoir_functional}
\end{align}
where $R: \mathcal{U}^h\to\R^N $ is the (finite-time) reservoir functional \cite{Grigoryeva2018}.
For most practical systems, it is often very difficult to find a closed form of the reservoir functional in Eq.~\eqref{eq:reservoir_functional}. 
But importantly, under these conditions the dynamical system is a causal, time-shift-invariant filter \cite{Gonon2020, Gonon2020a}. 
Incidentally, this also then guarantees the ergodicity of $r[k]$, as long as the input $u[k]$ is also ergodic.
For multi-dimensional dynamical systems ($N>1$), we assume that each component $r_n$ of the state vector $r$ is square-integrable with respect to the probability measure $\mu_u$ induced by the distribution of the input history, so that $r_n\in\mathcal{H}_u$.  

Now, to cleanly measure how much the dynamical system itself can approximate a target function, we cannot allow  arbitrary post-processing of $r[k]$. Directly using $r[k]$ would imply a computing system that cannot be tuned for different tasks, whereas allowing nonlinear transformations of $r[k]$ will make it difficult to determine which computation occurs due to the nonlinear dynamics of the system and which due to the post-processing.
We therefore allow a minimally adjustable, plain extraction of the information contained in $r[k]$, by introducing a trainable linear read-out that maps the system state vector to a scalar output $\tilde{z}\in\R$ as
\begin{align}
    \tilde{z} := \sum_{n=1}^Nw_nr_n[k] = w^{\mathsf{T}}r[k], 
    \label{eq:tildez}
\end{align}
where $w = (w_1, \ldots,w_N)^{\mathsf{T}}\in\R^{N}$ denotes the vector of linear read-out weights. This corresponds to the standard reservoir computing framework. 
To account for a constant offset, a practical approach is to include a bias term $w_0$ in the system output $\tilde{z}$ in Eq.~\eqref{eq:tildez}, yielding an affine instead of a linear read-out layer. 
Equivalently, we can assume that the reservoir states are centered,
\begin{align}
    \langle r_n\rangle_u = 0. 
\end{align} 
Accordingly, we will restrict our attention to target functions $z$ with zero mean for the rest of this manuscript, i.e., $\langle z\rangle_u = 0$. 

To evaluate the reconstruction accuracy of the system output $\tilde{z}$ relative to a target $z$, we simply use the mean squared error (MSE), 
\begin{align}
     &\mathrm{MSE}(z\mid r) \notag\\&:= \left\langle\left(\tilde{z} - z\right)^2\right\rangle_u\notag\\
    &= w^{\mathsf{T}}\langle rr^{\mathsf{T}}\rangle_uw  - 2w^{\mathsf{T}}\langle zr\rangle_u + \langle z^2\rangle_u. 
\end{align}
Notably, we extend the notation $\langle\cdot\rangle_u$ to vector- and matrix-valued functions by applying it element-wise, i.e., 
\begin{align}
    \left(\langle zr\rangle_u\right)_n &= \int_{\mathcal{U}^h} z(\omega)r_n(\omega)d\mu_u(\omega), \\
    \left(\langle rr^{\mathsf{T}}\rangle_u\right)_{mn}&= \int_{\mathcal{U}^h} r_m(\omega)r_n(\omega)d\mu_u(\omega).  
\end{align}
Minimizing this MSE with respect to the linear weights $w$ yields a normalized measure of how well the system can represent $z$, known as the information processing capacity (IPC)~\cite{Dambre2012}. 
\begin{align}
    C(z\mid r) := 1 - \frac{\min_w\mathrm{MSE}(z\mid r)}{\langle z^2\rangle_u}.
\end{align}
An IPC value of $1$ indicates perfect reconstruction of the target $z$, whereas an IPC value of $0$ means that the state implements no computational aspect of $z$. 
For later analysis, it is useful to write the IPC in an explicit closed form. 
From the first-order optimality condition, 
\begin{align}
    \pd{}{w}\mathrm{MSE}(z\mid r) &= 2\langle rr^{\mathsf{T}}\rangle_uw - 2 \langle zr\rangle_u = 0,
\end{align}
the optimal linear weight $w^*$ is 
\begin{align}
    w^*:= (\langle rr^{\mathsf{T}}\rangle_u)^{\dagger}\langle zr\rangle_u,
\end{align}
where $(\cdot)^{\dagger}$ denotes the Moore-Penrose pseudoinverse. 
Substituting this into the IPC formulation yields 
\begin{align}
    C(z\mid r) = \frac{\langle zr\rangle_u^{\mathsf{T}}\left(\langle rr^{\mathsf{T}}\rangle_u\right)^{\dagger}\langle zr\rangle_u }{\langle z^2\rangle_u}. \label{eq:czr}
\end{align}

Let $\{y_l\}_{l=1}^{\infty}\subset\mathcal{H}_u$ be a complete orthonormal basis of this space. 
Summing the IPCs over this basis allows us to quantify the overall computational capacity of the dynamical system without double-counting overlapping target functions. 
In particular, when the system state is full rank, we obtain  
\begin{align}
    \sum_{l=1}^L C(y_l\mid r) \le N. \label{eq:totalipc}
\end{align}
Intuitively, the attainment of the upper bound $N$ means that each node contributes along a linearly independent direction, fully utilizing the system's degrees of freedom. 
This key result from~\cite{Dambre2012} implies that the number of observed system dimensions inherently limits the total number of independent target functions that can be accurately reconstructed. 

\subsection{Noise-degraded information processing capacity \label{subsec:ipcundernoise}} 
In practical physical substrates, noise is inevitably intertwined with the system's dynamics. 
The presence of noise contaminates the information flow, leading to a degradation in the quality of state representations. 
This effect can be quantified as a reduction in the IPC.
Specifically, Polloreno~\textit{et al.} \cite{Polloreno2023} demonstrated that the total IPC decreases as a function of the noise-to-signal ratio. 
In this section, we introduce a model of dynamical systems with noise that is equivalent to the framework of~\cite{Polloreno2023}. 
However, we modify the deduction in a way that resembles the Hilbert space perspective of Kubota~\textit{et al.} \cite{Kubota2021}.
Here, our goal is to explicitly include all the different types of noise that dynamical systems may be subject to (input, process and output noise, additive or otherwise) and to later distinguish unperturbed dynamics from the noise-free part of stochastic dynamics obtainable via ensemble averages.

To explicitly describe the dependence of the system state on both the input and the noise, we introduce a noise source $v[k]\in\mathcal{V}$ and define its history as $v^{(h)}[k] = (v[k], \ldots, v[k-h+1])\in\mathcal{V}^h$~\cite{Kubota2021}. 
Analogously to $\mathcal{H}_u$, we introduce the Hilbert space 
\begin{align}
    \mathcal{H}_v := \{\zeta: \mathcal{V}^h\to\R\mid\langle\zeta^2\rangle_v<\infty\},
\end{align}
where $\langle\cdot\rangle_v$ denotes expectation with respect to the probability measure induced by the noise history. 

For measurable functions $\Omega: \mathcal{U}^h\times\mathcal{V}^h\to\R$, we define the expectation
\begin{align}
    \langle \Omega\rangle_{u,v} := \int_{\mathcal{U}^h\times\mathcal{V}^h}\Omega(\omega_u, \omega_v)d(\mu_u\otimes\mu_v)(\omega_u,\omega_v), 
\end{align}
where $\omega_u\in\mathcal{U}^h$ and $\omega_v\in\mathcal{V}^h$ denote realizations of the input and noise histories, respectively. 
We assume that the noise $v$ does not depend on the input $u$.
We then introduce the Hilbert tensor product 
\begin{align}
    \mathcal{H}_{u,v}:=\mathcal{H}_u\otimes \mathcal{H}_v, 
\end{align}
equipped with the inner product 
\begin{align}
    \langle \Omega, \eta\rangle_{u,v} := \langle \Omega\eta\rangle_{u,v},
\end{align}
and the induced norm 
\begin{align}
    \|\Omega\|_{u,v} := \sqrt{\langle \Omega, \Omega\rangle_{u,v}} = \sqrt{\langle\Omega^2\rangle_{u,v}}. 
\end{align}

We denote the observed state with noise by $r'[k] := (r_1'[k], \ldots, r_N'[k])^{\mathsf{T}}\in\R^N$, and regard the observed state as being determined by the truncated histories of both the input and the noise, 
\begin{align}
    r'[k] = R'(u^{(h)}[k], v^{(h)}[k]), \quad R': \mathcal{U}^h\times\mathcal{V}^h\to\R^N, 
\end{align}
where $R'$ is the reservoir functional for the observed state. 
Each component satisfies $r'_n\in\mathcal{H}_{u,v}$. 
Here, for notational simplicity, we denote the input and noise history $u^{(h)}[k]$ and $v^{(h)}[k]$ simply by $u$ and $v$, respectively. 

We define the closed subspace
\begin{align}
    \mathcal{M}_{u} := \mathcal{H}_u\otimes \mathrm{span}\{1\} \subset\mathcal{H}_{u,v},
\end{align}
which consists of all functions in $\mathcal{H}_{u,v}$ that depend only on the input history, where $1(v)\equiv 1$. 
Since $\mathcal{M}_{u}$ is a closed subspace of the Hilbert space $\mathcal{H}_{u,v}$, each component of $r'(u,v)$ admits the unique orthogonal decomposition
\begin{align}
    r'_n(u,v) = (\bar{r}_n\otimes1)(u,v) + \xi_n(u,v), 
\end{align}
where $\bar{r}_n\in\mathcal{H}_u$ is the unique function such that $\bar{r}_n\otimes 1$ is the orthogonal projection of $r_n'$ onto $\mathcal{M}_{u}$, and $\xi_n\in\mathcal{M}_{u}^{\perp}$ is the residual fluctuation term. 
By a slight abuse of notation, we identify each function $z\in\mathcal{H}_u$ with its embedding $z\otimes 1\in\mathcal{M}_u$, and hence write $z(u)$ in place of $(z\otimes 1)(u,v)$ whenever no confusion arises. 
Thus, the noise-free component $\bar{r}_n(u)$ also corresponds to the average over $v$, i.e, the projection of $r'_n$ onto $\mathcal{M}_{u}$:
\begin{align}
    \bar{r}_n(u) = \langle r'_n(u, v) \rangle _v \label{eq:r_bar}
\end{align}

With this convention, the orthogonal decomposition is simply written as 
\begin{align}
    r_n'(u,v) = \bar{r}_n(u) + \xi_n(u,v), 
\end{align}
and in vector notation, 
\begin{align}
    r'(u,v) = \bar{r}(u) + \xi(u,v). 
\end{align}
Since each component of the perturbation vector $\xi(u,v)$ belongs to $\mathcal{M}_{u}^{\perp}$, it is orthogonal to any noise-independent function, i.e., every element of $\mathcal{M}_{u}$.

Therefore, for any function $z\in\mathcal{H}_u$, the cross-correlation satisfies 
\begin{align}
    \langle z(u)\xi(u,v)\rangle_{u,v} = 0. 
\end{align}
Consequently, the cross-correlation satisfies, 
\begin{align}
    &\langle z(u)r'(u,v)\rangle_{u,v} \notag\\
    &= \langle z(\bar{r} + \xi)\rangle_{u,v}\notag\\
    &= \langle z\bar{r}\rangle_{u,v} + \langle z\xi\rangle_{u,v}\notag\\
    &= \langle z\bar{r}\rangle_u,
    \label{eq:zr'}
\end{align}
and similarly, the covariance of the observed state expands as
\begin{align}
    &\langle r'r'^{\mathsf{T}}\rangle_{u,v} \notag\\&= \langle (\bar{r} + \xi)(\bar{r} + \xi)^{\mathsf{T}}\rangle_{u,v}\notag\\
    &= \langle \bar{r}\bar{r}^{\mathsf{T}}\rangle_{u,v} + \langle\bar{r}\xi^{\mathsf{T}}\rangle_{u,v} + \langle \xi\bar{r}^{\mathsf{T}}\rangle_{u,v} + \langle \xi\xi^{\mathsf{T}}\rangle_{u,v}\notag\\
    &= \langle\bar{r}\bar{r}^{\mathsf{T}}\rangle_u + \langle\xi\xi^{\mathsf{T}}\rangle_{u,v}. 
\end{align}

Using these settings, the IPC of the noisy state $r'(u,v)$ for a target function $z$ can be derived as 
\begin{align}
    C(z\mid r')    &= \frac{\langle zr'\rangle_{u,v}^{\mathsf{T}}\left(\langle r'r'^{\mathsf{T}}\rangle_{u,v}\right)^{\dagger}\langle zr'\rangle_{u,v} }{\langle z^2\rangle_u}\notag\\
    &= \frac{\langle z\bar{r}\rangle_u^{\mathsf{T}}\left(\langle\bar{r}\bar{r}^{\mathsf{T}}\rangle_u + \langle\xi\xi^{\mathsf{T}}\rangle_{u,v}\right)^{\dagger}\langle z\bar{r}\rangle_u }{\langle z^2\rangle_u}. \label{eq:ipcundernoise}
\end{align}
To quantify the total information processing capability without redundant contributions, we consider the sum of the IPCs over a complete orthonormal basis $\{y_l\}_{l=1}^{\infty}\subset\mathcal{H}_u$ in a manner as before. 
\begin{align}
    &\sum_{l=1}^LC(y_l\mid r') \notag\\&= \sum_{l=1}^L\langle y_l\bar{r}\rangle_u^{\mathsf{T}}\left(\langle\bar{r}\bar{r}^{\mathsf{T}}\rangle_u + \langle \xi\xi^{\mathsf{T}}\rangle_{u,v}\right)^{\dagger}\langle y_l\bar{r}\rangle_u.
\end{align}
We introduce a linear transformation to orthonormalize the noise-free component $\bar{r}$. 
Let the eigen-decomposition of the covariance matrix of the noise-free component be $\langle\bar{r}\bar{r}^{\mathsf{T}}\rangle_u =: \bar{V}\bar{\Lambda}^2\bar{V}^{\mathsf{T}}$, where $\bar{\Lambda}^2\in\R^{d_{\bar{r}}\times d_{\bar{r}}}$ is a diagonal matrix of the $d_{\bar{r}}$ non-zero eigenvalues (i.e., $d_{\bar{r}} = \mathrm{rank}(\langle\bar{r}\bar{r}^{\mathsf{T}}\rangle_u)$), and $\bar{V}\in\R^{N\times d_{\bar{r}}}$ is the matrix of the corresponding orthonormal eigenvectors. 
We define the orthonormalized noise-free component $\hat{\bar{r}}\in\R^{d_{\bar{r}}}$ as 
\begin{align}
    \hat{\bar{r}}(u) := \bar{\Lambda}^{-1}\bar{V}^{\mathsf{T}}\bar{r}(u). 
\end{align}
Substituting $\bar{r} = \bar{V}\bar{\Lambda}\hat{\bar{r}}$ into the expression of the IPC gives 
\begin{align}
    &C(y_l\mid r') \notag\\
    &= \langle y_l\hat{\bar{r}}\rangle_u^{\mathsf{T}}\left(\langle\hat{\bar{r}}\hat{\bar{r}}^{\mathsf{T}}\rangle_u + \bar{\Lambda}^{-1}\bar{V}^{\mathsf{T}}\langle\xi\xi^{\mathsf{T}}\rangle_{u,v}\bar{V}\bar{\Lambda}^{-1}\right)^{-1}\langle y_l\hat{\bar{r}}\rangle_u\notag\\
    &= \langle y_l\hat{\bar{r}}\rangle_u^{\mathsf{T}}(I_{d_{\bar{r}}} + \hat{Q})^{-1}\langle y_l\hat{\bar{r}}\rangle_u,
\end{align}
where $I_{d_{\bar{r}}}$ is the $d_{\bar{r}}\times d_{\bar{r}}$ identity matrix, and $\hat{Q} = \bar{\Lambda}^{-1}\bar{V}^{\mathsf{T}}\langle\xi\xi^{\mathsf{T}}\rangle_{u,v}\bar{V}\bar{\Lambda}^{-1}\in\R^{d_{\bar{r}}\times d_{\bar{r}}}$ represents the noise covariance matrix normalized by the noise-free component variance. 
This term effectively corresponds to the noise-to-signal ratio in the feature space. 
Thus, the total IPC is derived as 
\begin{align}
    &\sum_{l=1}^LC(y_l\mid r') \notag\\
    &= \sum_{l=1}^L\langle y_l\hat{\bar{r}}\rangle_u^{\mathsf{T}}(I_{d_{\bar{r}}} + \hat{Q})^{-1}\langle y_l\hat{\bar{r}}\rangle_u\notag\\
    &= \sum_{l=1}^L\mathrm{Tr}\left[(I_{d_{\bar{r}}} + \hat{Q})^{-1}\langle y_l\hat{\bar{r}}\rangle_u\langle y_l\hat{\bar{r}}\rangle_u^{\mathsf{T}}\right]\notag\\
    &\le \mathrm{Tr}\left[(I_{d_{\bar{r}}} + \hat{Q})^{-1}\sum_{l=1}^{\infty}\langle y_l\hat{\bar{r}}\rangle_u\langle y_l\hat{\bar{r}}\rangle_u^{\mathsf{T}}\right]\notag\\
    &= \mathrm{Tr}[(I_{d_{\bar{r}}} + \hat{Q})^{-1}I_{d_{\bar{r}}}]\notag\\
    &= \sum_{n=1}^{d_{\bar{r}}} \frac{1}{1 + \hat{q}_n}, 
\end{align}
where $\{\hat{q}_n\}_{n=1}^{d_{\bar{r}}}$ is the set of eigenvalues of the normalized noise covariance $\hat{Q}$. 
The completeness of the basis $\{y_l\}_{l=1}^{\infty}$ ensures that the infinite sum recovers the covariance matrix of the orthonormalized state, $\langle \hat{\bar{r}}\hat{\bar{r}}^{\mathsf{T}}\rangle_u = I_{d_{\bar{r}}}$. 
In particular, when the noise-free component is full rank,
\begin{align}
    \sum_{l=1}^LC(y_l\mid r') \le\sum_{n=1}^N\frac{1}{1 + \hat{q}_n}. \label{eq:totalipcnoise}
\end{align}
Equation~\eqref{eq:totalipcnoise} was first derived by Hu \textit{et al.} \cite{hu2023tackling} and independently proposed by Polloreno \cite{Polloreno2023}.
The upper bound is achieved when the noise-free component has full rank and lies entirely in the span of the basis functions $\{y_l\}_{l=1}^L$. 
This result demonstrates that the total IPC monotonically decreases as the eigenvalues of $\hat{Q}$ (representing the noise-to-signal ratio of each mode) increase. 
Conversely, in the noise-free situation where $\hat{q}_n = 0$ for all $n$, the expression reduces to the noise-free total IPC given in Eq.~\eqref{eq:totalipc}.

\subsection{Distinguishing unperturbed and noise-free dynamics\label{sec:r=barr}}

Here, we note that the unperturbed state $r$ and the noise-free component $\bar{r}$ do not, in general, coincide, which was not stated clearly in~\cite{Polloreno2023}. 
This discrepancy arises because $\bar{r}$ is obtained by projecting the observed state $r'$ onto the closed subspace $\mathcal{M}_u$, which merely removes the residual noise fluctuation $\xi$ orthogonal to $\mathcal{M}_u$ rather than reproducing the unperturbed dynamics $r$. 
In the following, we will show a minimal example where noise leads to the appearance of IPCs in the noise-free part $\bar{r}$, which were not present in the unperturbed system $r$.

As a simple conceptual example, consider a one-dimensional reservoir state without memory, with added input noise. 
We assume that the input $u[k]$ is uniformly distributed on $[-1, 1]$, and the Legendre polynomials therefore form a natural orthonormal basis of $\mathcal{H}_u$. 
We choose the reservoir dynamics directly described by the reservoir functional as 
\begin{align}
    r'(u[k],v[k]) &= r(u[k] + v[k]), \\
    r(u[k]) &= \frac{1}{2}(5u^3[k]-3u[k]) = \frac{1}{\sqrt{7}}P_3(u[k]), \label{eq:P3-reservoir}
\end{align}
where
\begin{align}
    P_3(u[k]) := \frac{\sqrt{7}}{2} (5u^3[k]-3u[k])
\end{align}
denotes the normalized third-order Legendre polynomial. 

Therefore, under unperturbed conditions, the reservoir has 
\begin{align}
    C(P_3(u[k])\mid r) = 1, 
\end{align}
and, for any $z\in\mathcal{H}_u$ orthogonal to $P_3(u[k])$, 
\begin{align}
    C(z\mid r) = 0. 
\end{align}

For notational simplicity, throughout the remainder of this section, $u$ and $v$ denote $u[k]$ and $v[k]$, respectively. 
We will now deduce the IPC of this system under additive Gaussian input noise.
We therefore define the noisy reservoir state as $r'(u,v) = r(u+v)$, yielding
\begin{align}
    r'(u,v) &= \frac{1}{2}(5(u+v)^3 - 3(u+v))\notag\\
    &= r(u) + \frac{15}{2}u^2v + \frac{15}{2}uv^2 + \frac{5}{2}v^3 - \frac{3}{2}v.
\end{align}
Here, $v$ is a zero-mean Gaussian random noise with standard deviation $\sigma$. 
Then, by Eq.~\eqref{eq:r_bar}, for this system we obtain:
\begin{align}
    \bar{r}(u) &= r(u) + \frac{15}{2} \sigma^2 u
\end{align}
The IPC of a one-dimensional reservoir can be written as 
\begin{align}
    C(z\mid r') = \frac{(\langle zr'\rangle_{u,v})^2}{\langle z^2\rangle_u\langle r'^2\rangle_{u,v}} . 
\end{align}
Evaluating the terms in the above expression for the first-order Legendre polynomial, $P_1(u) := \sqrt{3}u$, and the third-order Legendre polynomial $P_3(u)$ yields (see Appendix.~\ref{sec:r=barr_calc} for the detailed calculation)
\begin{align}
    C(P_1(u)\mid r') &= \frac{\frac{75}{4}\sigma^4}{\frac{1}{7} + 6\sigma^2 + \frac{285}{4}\sigma^4 + \frac{375}{4}\sigma^6 },\\
    C(P_3(u)\mid r') &= \frac{\frac{1}{7}}{\frac{1}{7} + 6\sigma^2 + \frac{285}{4}\sigma^4 + \frac{375}{4}\sigma^6 }. 
\end{align}
We can also calculate the IPC of the noise-free part itself and obtain:
\begin{align}
    C(P_1(u)\mid \bar{r}) &= \frac { \frac{75}{4} \sigma^4 }{ \frac{1}{7} + \frac{75}{4} \sigma^4 } \\
    C(P_3(u)\mid \bar{r}) &= \frac { \frac{1}{7} }{ \frac{1}{7} + \frac{75}{4} \sigma^4 }. 
\end{align}
And in particular it holds that
\begin{align}
    C(P_1(u)\mid \bar{r}) + C(P_3(u)\mid \bar{r}) &= 1
\end{align}
for all inputs $u$. 
For the noise-free component $\bar{r}$, we can see that noise has a redistributary effect and does neither destroy nor enhance total IPC. 
The noise-degradation can also be explicitly seen by reading:
\begin{align}
    C(z \mid r') = C(z \mid \bar{r} )\frac{ \langle \bar{r}^2\rangle_u }{\langle r'^2\rangle_{u,v}}
\end{align}
Where the IPC of the noisy state $r'$ factorizes into that of $C(z \mid \bar{r} )$, which contains redistribution, and a term measuring the decay of the signal strength.

\begin{figure}[htpb]
  \centering
  \includegraphics[width=0.43\textwidth]{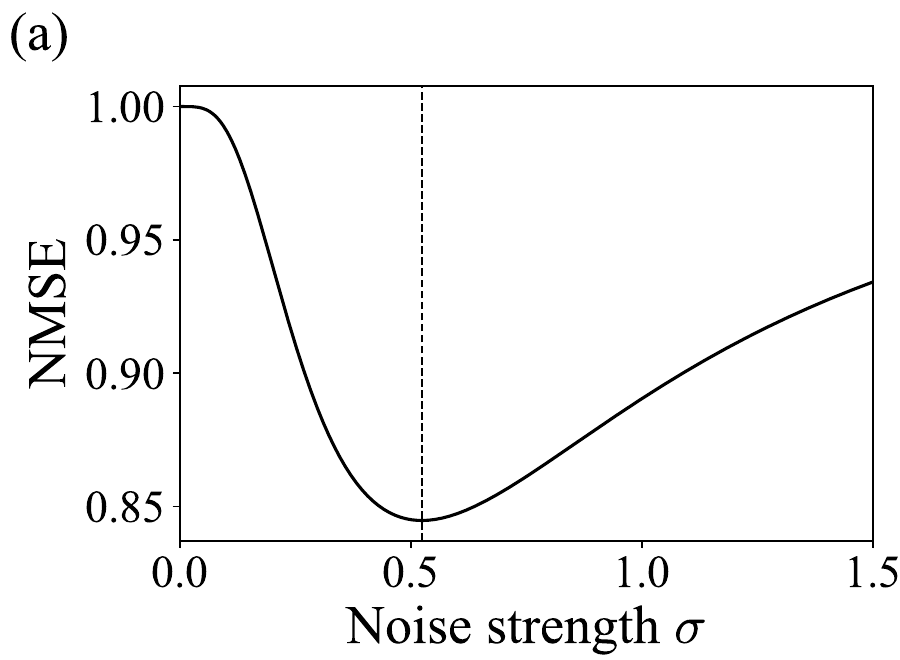}
  \includegraphics[width=0.43\textwidth]{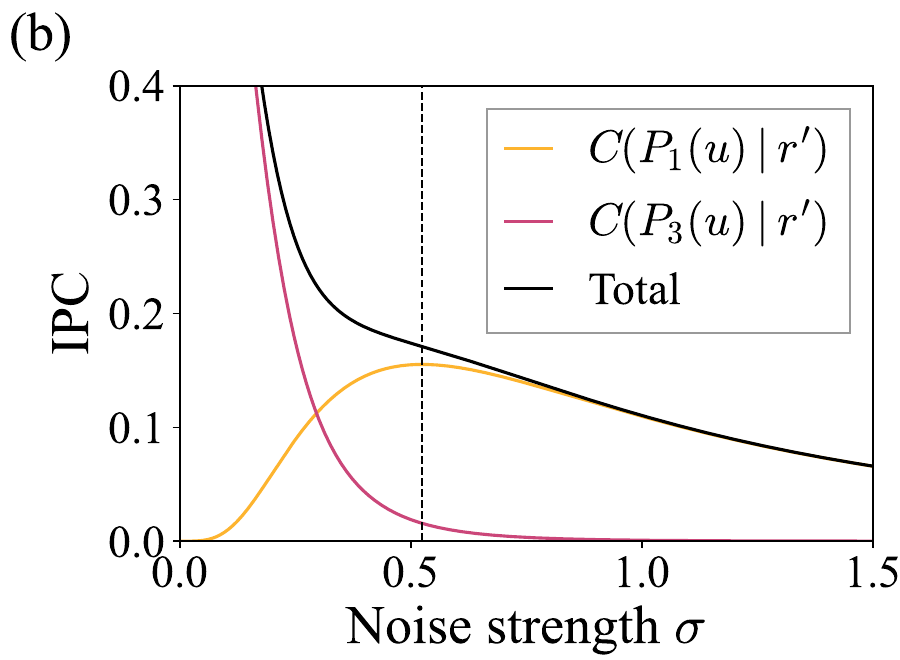}
  \caption{Noise-induced enhancement of information processing in a one-dimensional reservoir (Eq.~\eqref{eq:P3-reservoir}). (a) Normalized mean squared error (NMSE) for reconstructing the first-order Legendre polynomial $P_1(u)$ as a function of the noise strength $\sigma$. The minimum at intermediate noise strength shows that noise improves the reconstruction of this target. (b) Information processing capacities for the first- and third-order Legendre polynomials, $C(P_1\mid r')$ and $C(P_3\mid r')$, and their sum. While the capacity for $P_3$ decreases with increasing noise, the capacity for $P_1$ initially increases and reaches a maximum near $\sigma\approx0.5$, demonstrating that noise can enhance performance on a specific task even as the total capacity decreases.
}
  \label{fig:r=barr}
\end{figure}
Figure~\ref{fig:r=barr} shows how the dynamics of the one-dimensional reservoir are modified by the addition of noise, enabling information processing for previously inaccessible target functions. 
Figure~\ref{fig:r=barr}(a) shows the NMSE for reconstructing the first-order Legendre polynomial $P_1(u)$, which corresponds to $1 - C(P_1\mid r')$. 
Under unperturbed dynamics, the reservoir state contains only the $P_3(u)$ component and therefore has no information about $P_1(u)$, resulting in an NMSE of $1$. 
By introducing input noise, the NMSE is reduced, indicating that the reservoir acquires information about $P_1(u)$. 
The optimal noise strength is found to be approximately $\sigma = 0.5$. 
Figure~\ref{fig:r=barr} shows $C(P_1(u)\mid r')$, $C(P_3(u)\mid r')$, and their sum as functions of the noise strength $\sigma$. 
Under unperturbed conditions, $\sigma = 0$, the reservoir has $C(P_3(u) \mid r) = 1$ and $C(P_1(u)\mid r)=0$. 
As the noise strength $\sigma$ increases, $C(P_1(u)\mid r')$ increases, and reaches a maximum around $\sigma\approx 0.5$, consistent with the minimum NMSE in Fig.~\ref{fig:r=barr}(a). 
Ultimately, $C(P_1(u)\mid r')$ and the total IPC decreases for larger $\sigma$, since the reservoir state becomes dominated by noise. 

Such a non-monotonic behavior is similar to stochastic resonance~\cite{Wiesenfeld1995, Hanggi2002}, however, it does not represent a clean example of it: Here, the noise does not enhance a sub-threshold signal, but instead redistributes signals across different nonlinear representations.

\subsection{Estimation of noise-free information processing capacity via CROP\label{subsec:proposed}}
We now turn our attention to the practical application of measuring the IPC with noisy data.
When measuring the noise-degraded IPC $C(z \mid r')$ via noisy data $r'$ of an unknown physical system, it is often impossible to discern whether shortcomings in the observed IPC reflect the system's inherent computational capability $C(z \mid \bar{r})$ or a degraded value caused by noise-induced perturbations. 
While noise-mitigation techniques such as filtering and signal averaging are commonly employed, many physical systems possess intrinsic noise that is fundamentally difficult to eliminate. 
Furthermore, applying such techniques to a prototype can be costly and inefficient. 

To address this, we propose the Covariance Reconstruction by Orthogonal Projection (CROP) method to estimate the noise-free IPC directly from noise-contaminated observations.
In particular, here we assume that we do not necessarily have a finely tuned numerical model of the target dynamical system.
In this case predicting the redistributory effect discussed in Sec.~\ref{sec:r=barr} is out of scope in this work (although we suspect it might be possible in principle).
Instead, we aim to reconstruct the noise-free part $\bar{r}$ for a fixed noise strength and distribution.

As detailed in Sec.~\ref{subsec:ipcundernoise}, the noisy state can be decomposed as $r'(u,v) = \bar{r}(u) + \xi(u,v)$, where $\bar{r}(u)$ represents the noise-free component and $\xi(u,v)$ is the perturbation term. 
Based on this formulation, the noise-free IPC to be estimated is given by
\begin{align}
    C(z\mid\bar{r}) = \frac{\langle z\bar{r}\rangle_u^{\mathsf{T}} (\langle\bar{r}\bar{r}^{\mathsf{T}}\rangle_u)^{\dagger}\langle z\bar{r}\rangle_u}{\langle z^2\rangle_u}. 
\end{align}
Compared to this noise-free IPC, the noise-degraded IPC, $C(z\mid r')$ in Eq.~\eqref{eq:ipcundernoise}, differs by the addition of the positive semi-definite noise covariance matrix $\langle\xi\xi^{\mathsf{T}}\rangle_{u,v}$ into the noise-free component covariance $\langle\bar{r}\bar{r}^{\mathsf{T}}\rangle_u$, leading to the degradation of the IPC. 
Consequently, the problem of estimating the noise-free IPC from noisy observations $r'(u,v)$ reduces to estimating the noise-free component covariance $\langle\bar{r}\bar{r}^{\mathsf{T}}\rangle_u$, which cannot be achieved by simply computing the covariance of the observed state $\langle r'r'^{\mathsf{T}}\rangle_{u,v}$. 

To address this, we introduce the following estimate of the noise-free IPC, $\tilde{C}(z)$, based only on the noisy state $r'(u,v)$ as
\begin{align}
    \tilde{C}(z) &:= \frac{\langle zr'\rangle_{u,v}^{\mathsf{T}}\left(\sum_{l=1}^L\langle y_lr'\rangle_{u,v}\langle y_lr'\rangle_{u,v}^{\mathsf{T}}\right)^{\dagger} \langle zr'\rangle_{u,v}}{\langle z^2\rangle_u}.\label{eq:cpot}
\end{align}
where $\{y_l\}_{l=1}^{\infty}$ is a complete orthonormal basis of the Hilbert space $\mathcal{H}_u$ of target functions defined on the input history $u^{(h)}[k]$. 
To demonstrate that $\tilde{C}(z)$ effectively approximates the noise-free IPC $C(z\mid\bar{r})$, we substitute $r'(u,v) = \bar{r}(u) + \xi(u,v)$ into Eq.~\eqref{eq:cpot} and then simplify the resulting expression using Eq.~\eqref{eq:zr'}. 
The estimated noise-free IPC can be rewritten as 
\begin{align}
    \tilde{C}(z) = \frac{\langle z\bar{r}\rangle_u^{\mathsf{T}}\left(\sum_{l=1}^L\langle y_l\bar{r}\rangle_u\langle y_l\bar{r}\rangle_u^{\mathsf{T}}\right)^{\dagger}\langle z\bar{r}\rangle_u}{\langle z^2\rangle_u}.
\end{align}
Due to the completeness of the basis $\{y_l\}_{l=1}^{\infty}$, the matrix in parentheses represents a truncated approximation of the covariance of the noise-free component, 
\begin{align}
    \sum_{l=1}^L\langle y_l\bar{r}\rangle_u\langle y_l\bar{r}\rangle_u^{\mathsf{T}}\approx\langle \bar{r}\bar{r}^{\mathsf{T}}\rangle_u.
\end{align}
Thus, the estimated noise-free IPC reduces to
\begin{align}
    \tilde{C}(z)&\approx\frac{\langle z\bar{r}\rangle_u^{\mathsf{T}}(\langle \bar{r}\bar{r}^{\mathsf{T}}\rangle_u)^{\dagger}\langle z\bar{r}\rangle_u}{\langle z^2\rangle_u} \notag\\
    &= C(z\mid \bar{r}).
\end{align}
The approximation becomes an exact equality if the noise-free component $\bar{r}(u)$ lies entirely in the span of $\{y_l\}_{l=1}^{L}$. 
In particular, in the limit $L\to\infty$, Parseval's identity guarantees the equality for any square-integrable state. 
Heuristically, the condition that $\bar{r}(u)$ is in the subspace spanned by $\{y_l\}_{l=1}^L$ can often be well approximated even with a finite basis set, provided that it covers a sufficiently long input history and a sufficiently rich class of functions. 

Due to the strong dependence of $\tilde{C}(z)$ on the reservoir architecture, a theoretical evaluation of the approximation error with respect to $L$ is difficult in general. 
We therefore rely on numerical validation in the following sections. 
In the numerical simulations, we use $L = 8902$ for the example of a classical reservoir, but this choice is not fundamental; it is chosen to ensure sufficient coverage of the relevant function space in practice. 
Note that increasing $L$ too much eventually degrades estimation performance as well. 

This method admits a natural interpretation in terms of the product space $\mathcal{H}_{u,v} =\mathcal{H}_u\otimes\mathcal{H}_v$. 
As discussed in Sec.~\ref{subsec:ipcundernoise}, the noise-free component $(\bar{r}\otimes 1)(u,v)$ belongs to the subspace $\mathcal{M}_u=\mathcal{H}_u\otimes\mathrm{span}\{1\}$, whereas the fluctuation term $\xi(u,v)$ belongs to $\mathcal{M}_u^{\perp}$. 
Consequently, the correlations $\langle y_lr'\rangle_{u,v}$ extract only the component of $r'(u,v)$ lying in $\mathcal{M}_u$, while an orthogonal fluctuation component does not contribute. 
The matrix $\sum_l \langle y_lr'\rangle_{u,v}\langle y_lr'\rangle_{u,v}^{\mathsf{T}}$ therefore reconstructs the covariance of the projected state $\bar{r}(u)$ from noisy observations. 
Since the reconstruction of the covariance of the noise-free component forms the core of the proposed noise-free IPC estimation method, we term the method covariance reconstruction by orthogonal projection (CROP). 


\section{Application to classical reservoirs}
\label{sec:classical}

\subsection{System definition and baseline unperturbed IPC\label{subsec:reconst}}%
In Sec.~\ref{subsec:proposed}, we proposed the CROP method to estimate the noise-free IPC from noise-contaminated observations, mitigating the effects of noise-induced perturbations. 
In this section, we apply this approach to numerical simulations of specific systems to verify its effectiveness. 

As a simple example of a discrete-time dynamical system, we consider a system with a quadratic nonlinearity function, evolving according to 
\begin{align}
    r[k] &= Ar[k-1] - (r[k-1]\odot r[k-1]) \notag\\
    &\quad+ Bu[k] + \sigma v[k], 
\end{align}
where $r[k]\in\R^N$ is the system state, $u[k]\in\R$ is the one-dimensional input, $v[k]\in\R^N$ is an additive noise, and $\sigma\ge 0$ controls the noise amplitude. 
$A\in\R^{N\times N}$ and $B\in\R^N$ are the state and input matrices, randomly set with scales ensuring that $r[k]$ does not diverge over time. 
The input $u[k]$ is assumed to be i.i.d., which allows a systematic decomposition of the input history without introducing correlations. 

As a baseline, we first consider the unperturbed case $(\sigma = 0)$, where the IPC can be computed analytically from the Wiener-Volterra series expansion. 
The target functions are Legendre polynomials of the input and their products at different time steps, e.g., $u[k-2]$, $u[k]u[k-3]$, and $\frac{1}{2}(3u[k-1]^2-1)$. 
These polynomials are orthogonal with respect to the uniform distribution on $[-1,1]$, which ensures consistency when we later use finite-length input series drawn from this distribution to compute the IPC. 
The method for solving this system analytically and recursively to obtain the Wiener-Volterra series is described in Appendix~\ref{sec:analytical}. 
Kubota et al.~\cite{Kubota2021} have shown that the Wiener-Volterra series is essentially equivalent to the IPC, and the procedure for converting it to the IPC is explained in Appendix~\ref{sec:volterra}. 

\begin{figure}[htpb]
  \centering
  \includegraphics[width=0.43\textwidth]{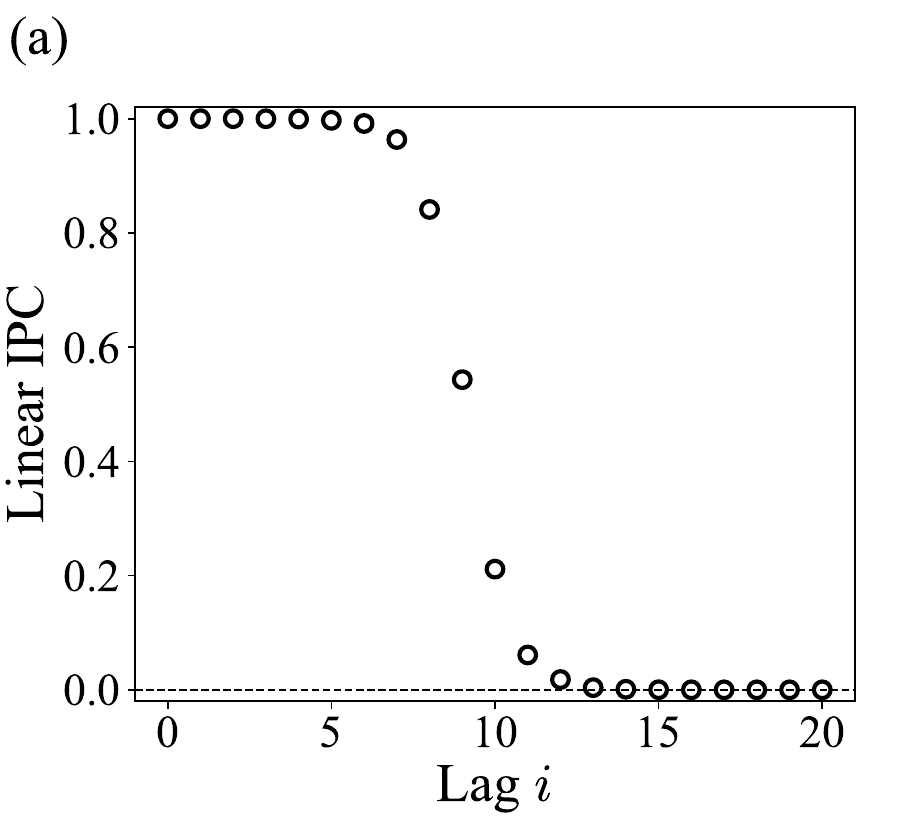}
  \includegraphics[width=0.43\textwidth]{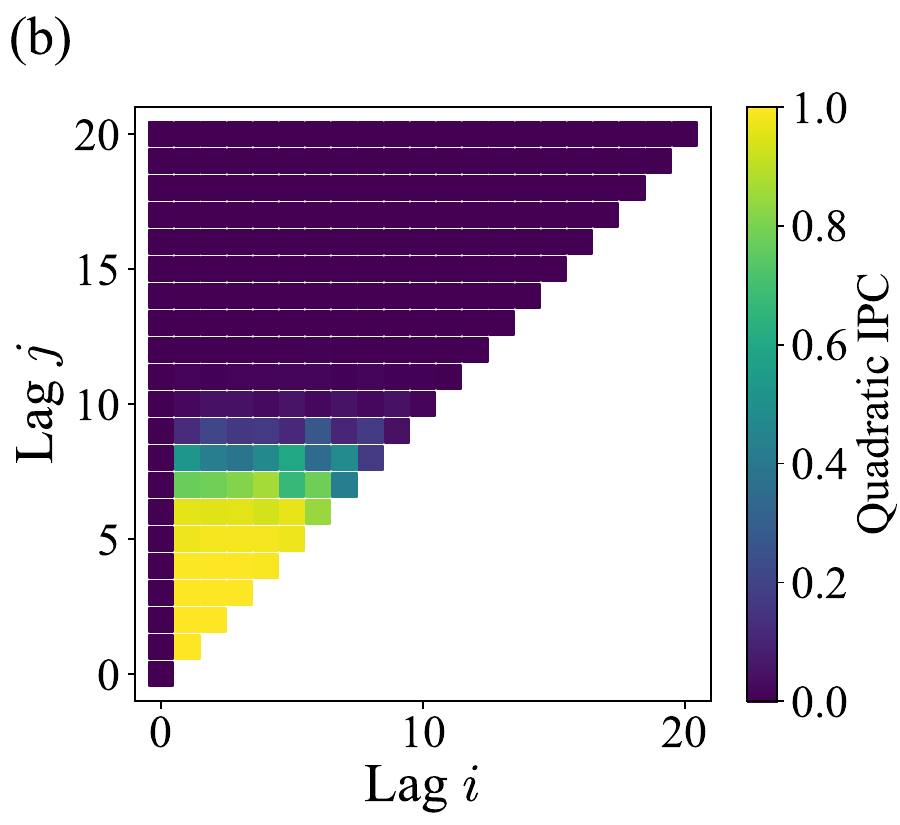}
  \caption{The IPC of a 100-node system with a quadratic nonlinearity in the unperturbed case. (a) The IPC for linear target functions, $u[k-i]$. (b) The IPC for quadratic target functions, $u[k-i]u[k-j]$ (off-diagonal) and $\frac{1}{2}(3(u[k-i])^2 - 1)$ (diagonal).}
  \label{fig:quad_analytical}
\end{figure}
Figure~\ref{fig:quad_analytical} shows the baseline IPC for $N = 100$ nodes in the unperturbed case. 
Figure~\ref{fig:quad_analytical}(a) shows the IPC for linear target functions, corresponding to $u[k], u[k-1], \ldots,$ with the horizontal axis representing the lag $i$ of $u[k-i]$. 
The IPC for $u[k]$ is close to its maximum value of $1$, and it decreases as the lag increase. 
This reflects the fact that in a dissipative system more recent inputs are retained with higher accuracy, while information from the more distant past gradually fades. 
Figure~\ref{fig:quad_analytical}(b) shows the IPC for quadratic target functions, visualized using a colormap. 
Off-diagonal elements correspond to products of inputs at different lags, with the $x$- and $y$-axes indicating the respective lags $i$ and $j$ for $u[k-i]u[k-j]$. 
Diagonal elements correspond to second-order Legendre polynomials of individual inputs, i.e., $\frac{1}{2}(3(u[k-i])^2 - 1)$. 
By construction, no quadratic term of the current input $u[k]$ appears, so the corresponding entries are zero. 
Otherwise, the trend is similar to the linear case: higher IPC values are associated with more recent input, which in the quadratic case appear in the lower-left region of the plot. 

The IPC can be computed for higher-order target functions in the same manner; however, the number of index combinations grows rapidly, making visualization in a two-dimensional plot impractical. 
In addition, higher-order terms with longer delays tend to have smaller contributions, as they involve repeated applications of dissipative $A$, leading to stronger attenuation. 
We compute the IPC exhaustively up to sixth-degree basis functions over index combinations with non-negligible contributions and aggregate them by polynomial order. 
While the aggregated IPC distributions shown later provide a useful summary of the overall structure, they can obscure the contributions of individual basis functions, making it important to keep in mind the delay-resolved components illustrated in Fig.~\ref{fig:quad_analytical}. 

\subsection{Estimation of the noise-free IPC from noise-contaminated time series}
Next, we compute the noise-degraded IPC from noisy time-series data and compare it with the analytical results. 
We then apply the proposed CROP method to estimate the noise-free IPC from the same data. 
To evaluate the IPC, we consider an i.i.d. input sequence $u[k]$ drawn from a uniform distribution over $[-1, 1]$, and generate a time series of the system state $r[k]$.
The noise term $v[k]\sim\mathcal{N}(0, I_N)$ has i.i.d. components, with $\sigma$ controlling the standard deviation. 

When evaluating the IPC from finite data, two approaches are naturally considered. 
One is to replace the expectation over $u$ and $v$ in Eq.~\eqref{eq:ipcundernoise} by empirical time averages. 
Another is a train/test protocol, in which the optimal weights $w^*$ are obtained from training data and the MSE is evaluated on separate test data. 
Both approaches introduce finite-sample bias in IPC evaluation, as discussed in Appendix~\ref{sec:samplebias}. 
In this paper, we adopt the train/test protocol to avoid overestimation of the IPC in finite samples. 
The same protocol is also used in the CROP framework. 
The definition of the estimated noise-free IPC differs from the simple time-average replacement of the expectation over $u$ and $v$ in Eq.~\eqref{eq:cpot}, and the resulting modified expression for $\tilde{C}(z)$ is detailed in Appendix~\ref{sec:traintest}. 

\begin{figure}[htpb]
  \centering
  \includegraphics[width=0.43\textwidth]{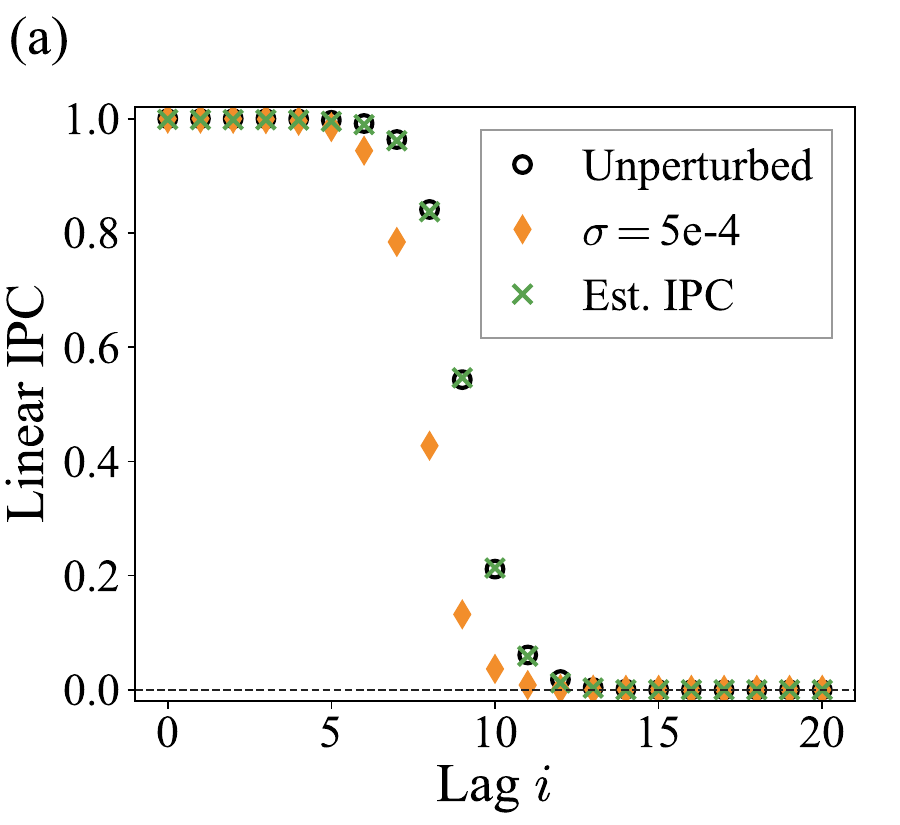}
  \includegraphics[width=0.43\textwidth]{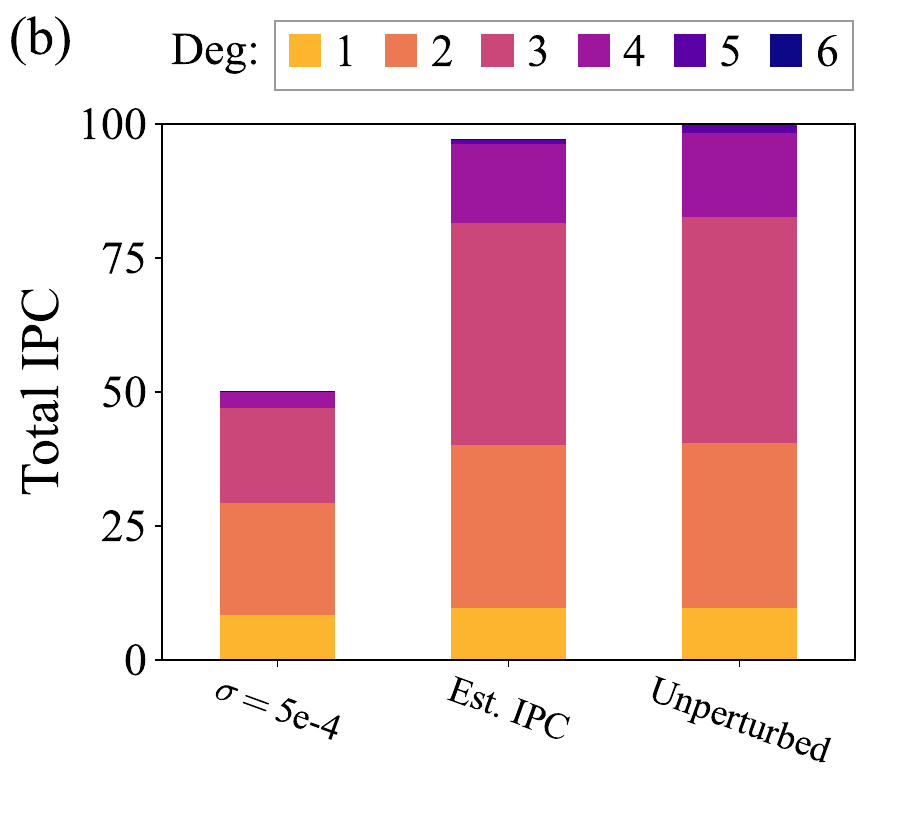}
  \caption{The IPC and the estimated noise-free IPC (`Est. IPC') of a 100-node system with quadratic nonlinearity under process noise. (a) The linear IPC $u[k-i]$: black circles, the unperturbed IPC (Fig.~\ref{fig:quad_analytical}(a)); orange diamonds, the IPC directly computed from noisy data ($\sigma = 5\times 10^{-4}$, $10^6$ data points); green crosses, the noise-free IPC estimated by the CROP. (b) The total IPC up to the sixth degree from the same noisy data as in (a), with bars (left to right) showing the noise-degraded IPC, the estimated noise-free IPC via CROP, and the unperturbed IPC. }
  \label{fig:quad_5e-4}
\end{figure}%
Figure~\ref{fig:quad_5e-4} shows the IPC computed directly from noisy time series data and the estimated noise-free IPC, using $10^6$ data points, with $5\times 10^5$ points allocated to the training set and the remaining $5\times 10^5$ points to the test set. 
In Fig.~\ref{fig:quad_5e-4}(a), the orange diamonds represent the noise-degraded linear IPC obtained with process noise of standard deviation $\sigma = 5\times 10^{-4}$, while the black circles denote the unperturbed IPC shown in Fig.~\ref{fig:quad_analytical}(a) for reference. 
While the IPCs at smaller lags decrease only slightly, those for $i = 6$ to $10$ exhibit a significant reduction. 
In contrast, the green crosses represent the noise-free IPC estimated from the same noisy time series, and they closely match the unperturbed baseline. 
Figure~\ref{fig:quad_5e-4}(b) shows bar charts of the total IPC for each polynomial order up to the sixth degree. 
From left to right, the bars correspond to the noise-degraded IPC, the estimated noise-free IPC via the proposed CROP method, and the unperturbed reference.  
When the IPC is computed directly in the presence of noise, the IPC values, particularly at the third and fourth degrees, are significantly reduced, dropping the total IPC to approximately 50. 
On the other hand, the estimated noise-free IPC largely matches the reference distribution, except for some higher-order components, indicating that CROP works effectively. 

\begin{figure}[htpb]
  \centering
  \includegraphics[width=0.44\textwidth]{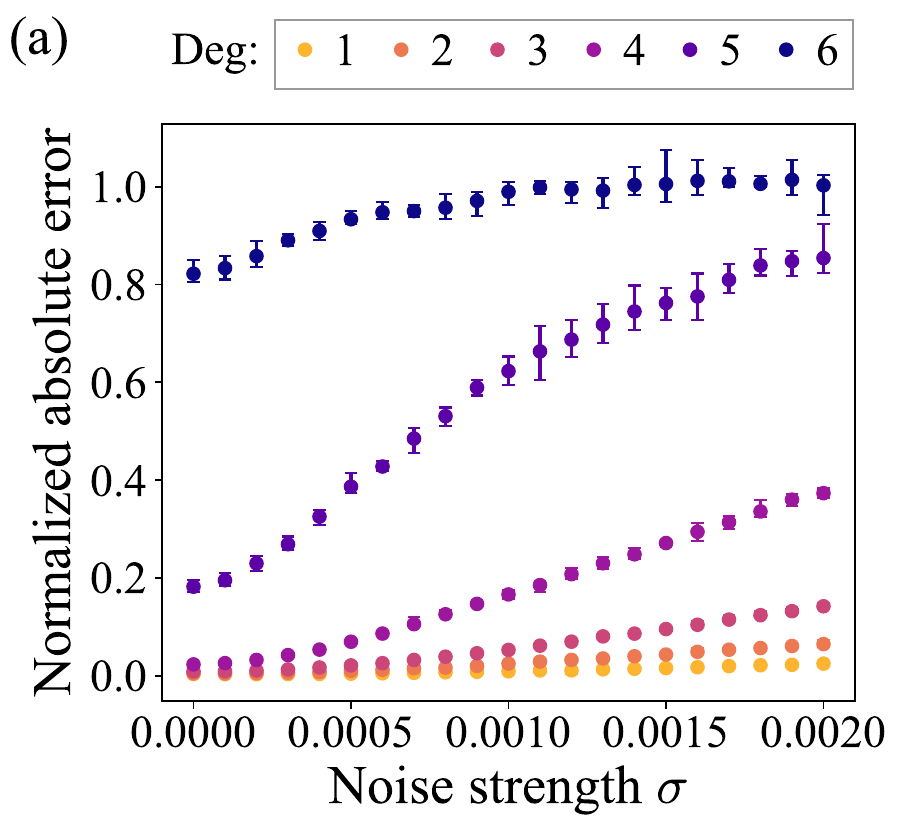}
  \includegraphics[width=0.44\textwidth]{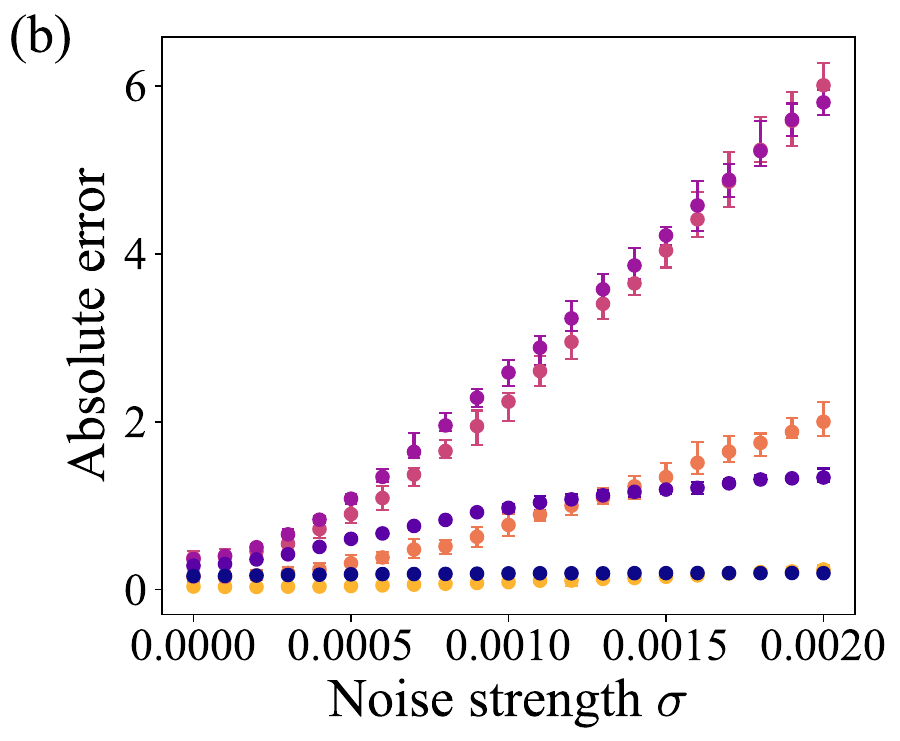}
  \caption{Evaluation of the noise-free IPC estimation of the CROP method in a 100-node system with quadratic nonlinearity under process noise. The estimation error is assessed for each polynomial order $d$. The markers represent averages over $10$ independent estimation trials using different input sequences of length $10^6$. (a) Normalized absolute error ($\mathrm{NAE}^{(d)}$), showing the relative deviation from the unperturbed IPC under various noise strengths $\sigma$. (b) Absolute error ($\mathrm{AE}^{(d)}$), showing the deviation from the unperturbed IPC under various noise strengths $\sigma$. }
  \label{fig:quad_AE}
\end{figure}
To evaluate the performance of CROP, we now quantify how close the estimated noise-free IPC is to the unperturbed IPC for each polynomial order $d$ by defining the normalized absolute error (NAE) as 
\begin{align}
    \mathrm{NAE}^{(d)} := \frac{\sum_{y_l\in\mathcal{Y}_d}|C(y_l\mid r) -\tilde{C}(y_l)|}{\sum_{y_l\in\mathcal{Y}_d}C(y_l\mid r)}, 
\end{align}
where $\{y_l\}$ are orthonormal functions in the Hilbert space $\mathcal{H}_u$ of target functions, and $\mathcal{Y}_d$ denotes the set of basis functions of degree $d$. 
$\mathrm{NAE}^{(d)}$ measures the relative deviation from the unperturbed IPC; if all $y_l\in\mathcal{Y}_d$ are estimated without error, such that $\tilde{C}(y_l) = C(y_l\mid r)$, then $\mathrm{NAE}^{(d)} = 0$. 
In contrast, if the estimation fails completely, i.e., the estimated noise-free IPC $\tilde{C}(y_l)$ is clipped to zero for all $y_l\in\mathcal{Y}_d$, then $\mathrm{NAE}^{(d)} = 1$. 
While the NAE measures the relative error for each polynomial order, it tends to overestimate the error for components with small IPC contributions. 
To complement this relative measure, we introduce the absolute error (AE), defined as 
\begin{align}
    \mathrm{AE}^{(d)} := \sum_{y_l\in\mathcal{Y}_d}|C(y_l\mid r) -\tilde{C}(y_l)|. 
\end{align}

Figures~\ref{fig:quad_AE}(a) and (b) show $\mathrm{NAE}^{(d)}$ and $\mathrm{AE}^{(d)}$ as functions of the noise strength $\sigma$, respectively. 
To assess the stability of the CROP, independent estimation trials are performed $10$ times using different input sequences, each consisting of $10^6$ samples. 
The plotted values represent the averages over the $10$ trials, while the error bars indicate the minimum and maximum values. 
From $\mathrm{NAE}^{(d)}$, it is clear that lower-order components are estimated more accurately, with degrees up to three remaining around $0.1$ even at relatively high noise levels, which demonstrates that most of their contribution is successfully estimated. 
This suggests that lower-order components are more redundantly encoded in the system state and thus less affected by noise, as also supported by Fig.~\ref{fig:quad_5e-4}(a), where the orange diamonds remain close to the unperturbed black circles at small lags. 
While the NAE alone implies that the higher-order components, specifically the fifth and sixth degrees, are poorly estimated, the AE plot reveals that their absolute errors are small, owing to their inherently small contributions in the unperturbed case. 
Furthermore, although the variability across trials becomes relatively larger for higher-order components, the corresponding error bars remain small in both the NAE and AE plots. 
These results indicate that the CROP provides stable estimates across different input realizations and effectively captures the dominant contributing components despite the presence of noise. 

\subsection{Comparison of the noise-free IPC estimation with conventional ensemble averaging}
So far, we have shown that the CROP effectively estimates the noise-free IPC by reconstructing the covariance of the noise-free component. 
However, as a simpler alternative, one can also improve the effective signal-to-noise ratio by averaging time series obtained from repeated trials under the same input sequence, as is commonly done in physical computing systems. 
Here, we examine whether our method outperforms this conventional approach. 

Let $K_{\text{total}}$ describe the total number of physical measurements of the system state. 
If part of this data is used for averaging, a natural tradeoff between noise-reduction and the size of the remaining time series $r'(k)$ emerges.
For example, by repeating the same experiment twice (with exactly the same $u(k)$ but different noise realizations) and averaging the system states, we can reduce the noise, but we also reduce the total length of our training data set to $K_{\text{total}}/2$.
Accordingly, increasing the number of repetitions improves the signal-to-noise ratio, whereas the length of the time series available for each IPC computation becomes shorter. 
Since the IPC calculation relies on time averaging, excessively short time series lead to inaccurate evaluation. 
In the extreme case of $10^6$ repetitions, for example, each IPC computation is based on only a single time point, making reliable estimation impossible. 
In the following, we compare our method with this approach using the same total number of data points ($K_{\text{total}} = 10^6$).

\begin{figure}[htbp]
  \centering
  \includegraphics[width=0.45\textwidth]{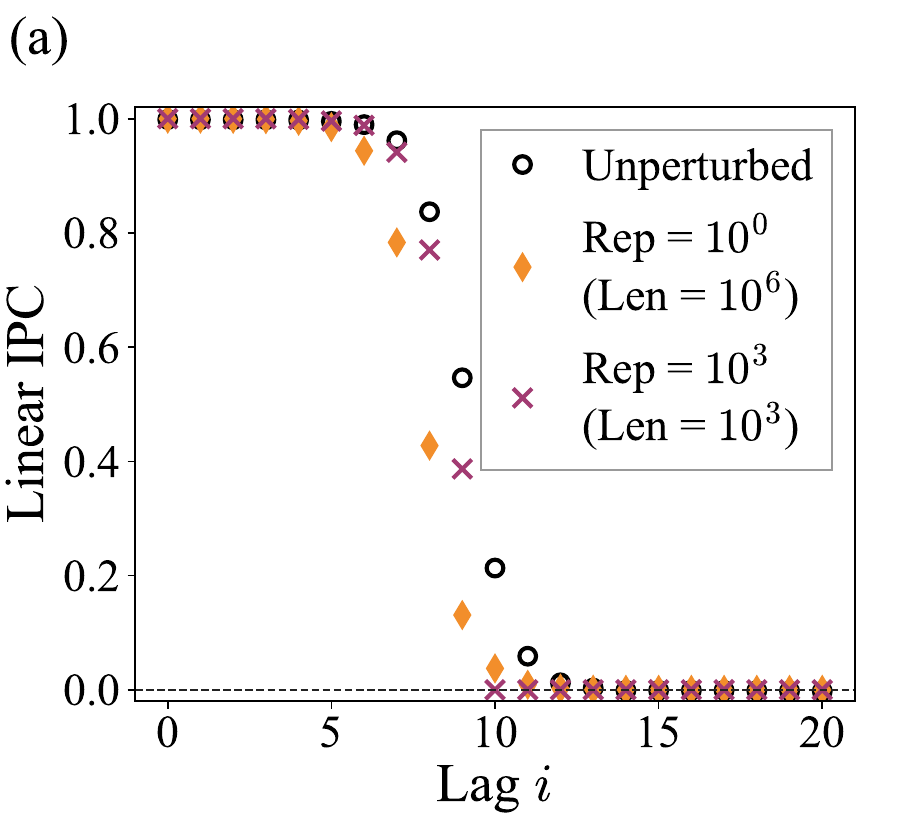}
  \includegraphics[width=0.45\textwidth]{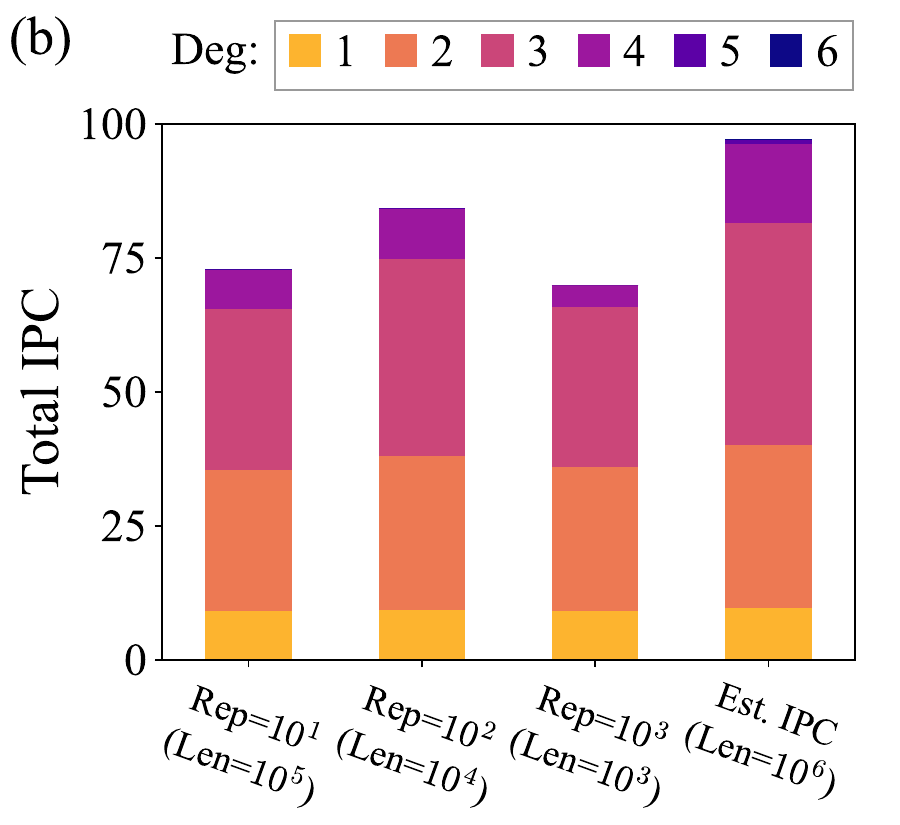}
  \caption{The IPC under repetition-based averaging for a 100-node system with quadratic nonlinearity under process noise. (a) The linear IPC $u[k-i]$: black circles, the unperturbed IPC; orange diamonds, the IPC computed from a single noisy time series ($10^6$ data points); purple crosses, the IPC obtained by averaging over $10^3$ repetitions (resulting in a time series of length $10^3$). (b) The total IPC up to the sixth degree, with bars (left to right) showing the IPC obtained from $10^1$, $10^2$ and $10^3$ repetitions (lengths are $10^5$, $10^4$ and $10^3$) via conventional averaging, and the estimated noise-free IPC via CROP for the same data. CROP outperforms averaging on sample efficiency.}
  \label{fig:quad_repeat}
\end{figure}
Figure~\ref{fig:quad_repeat} shows the IPC directly computed from time series obtained by averaging outputs over repeated trials with the same input sequence, under a noise standard deviation of $\sigma = 5\times 10^{-4}$. 
In Fig.~\ref{fig:quad_repeat}(a), the purple crosses represent the linear IPC computed after improving the signal-to-noise ratio by averaging over $10^3$ repetitions, resulting in a time series of length $10^3$. 
For reference, the orange diamonds correspond to the case of a single repetition (i.e., no averaging), where a time series of $10^6$ points is used, and the black circles denote the unperturbed IPC. 
The plot shows that for lags $i = 6$ to $9$, averaging over $10^3$ repetitions significantly mitigates the noise-induced reduction in the IPC, yielding values close to the unperturbed IPC even with this simple approach. 
However, for $i = 10$, the capacity obtained after $10^3$ repetitions is lower than that with a single repetition. 
This can be understood as a balance between the improvement in the calculated IPC due to the enhanced signal-to-noise ratio and the degradation caused by the shorter time series. 
Such loss of capacity due to shorter time series is more pronounced for higher-order basis functions, where the number of index combinations is large but each individual IPC is small. 
As shown in Fig.~\ref{fig:quad_repeat}(b), increasing the number of repetitions does not necessarily bring the total IPC up to the sixth degree closer to the unperturbed value of $100$; rather, while increasing repetitions from $10^1$ to $10^2$ (reducing the data length from $10^5$ to $10^4$) improves the total IPC, further increasing repetitions from $10^2$ to $10^3$ (reducing the data length from $10^4$ to $10^3$) results in a noticeable reduction. 
In contrast, the estimated noise-free IPC using the proposed CROP method (right most column in Fig.~\ref{fig:quad_repeat}(b)), calculated using $10^6$ data points without averaging, clearly exhibits a distribution closer to the unperturbed case (see Fig.~\ref{fig:quad_5e-4}). 

\begin{figure}[htpb]
  \centering
  \includegraphics[width=0.44\textwidth]{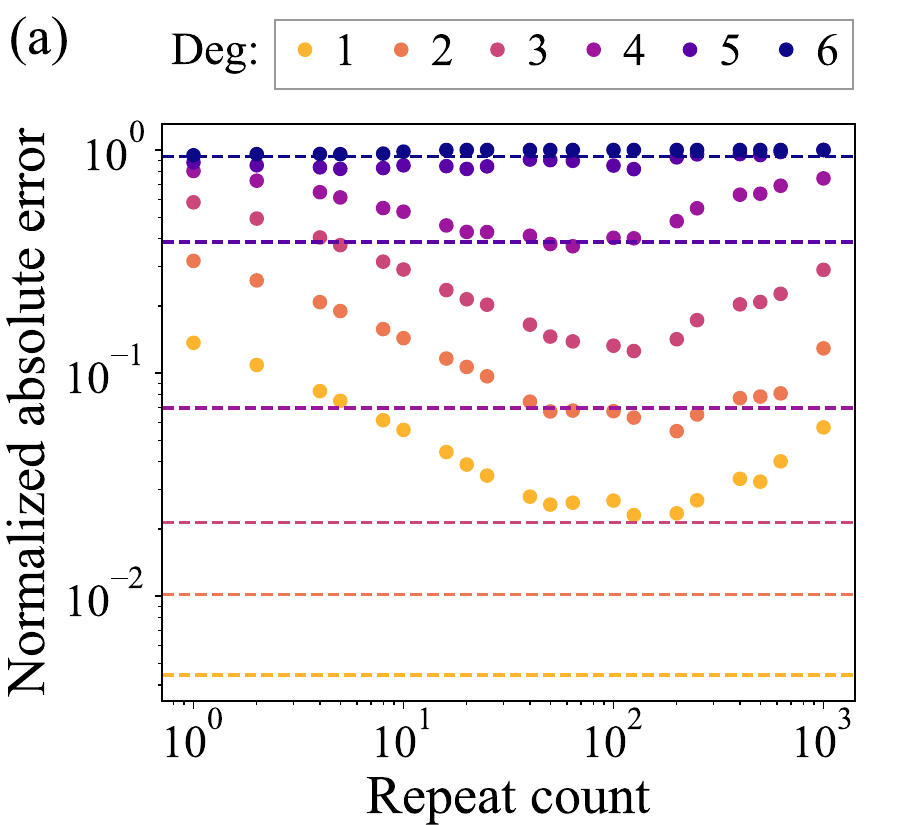}
  \includegraphics[width=0.44\textwidth]{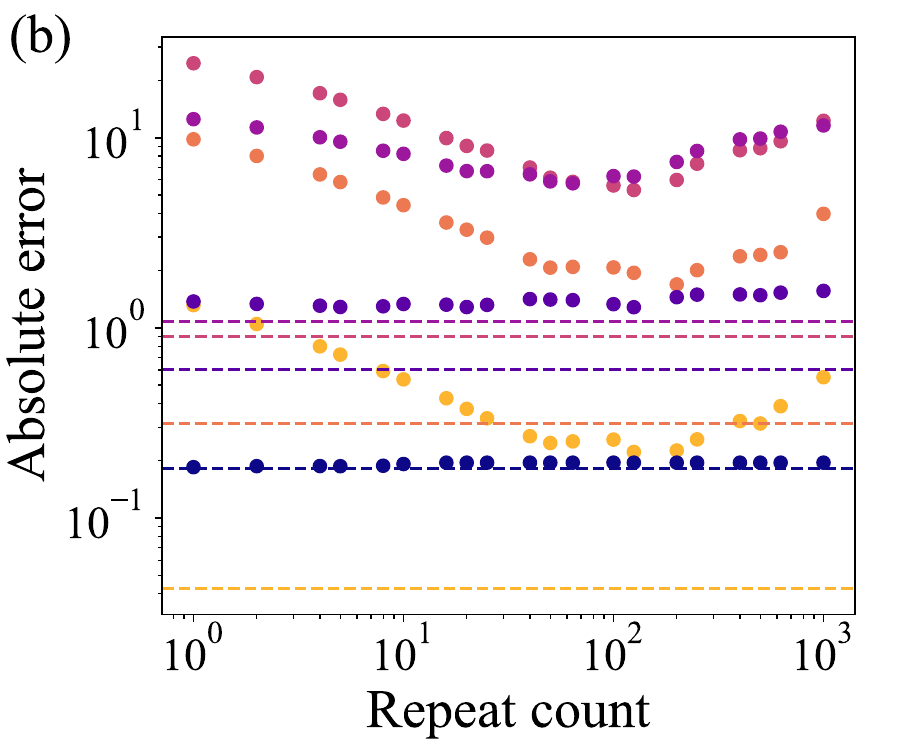}
  \caption{Evaluation of the IPC under repetition-based averaging (dots) in a 100-node system with quadratic nonlinearity under process noise. The IPC is compared with CROP method (dotted lines) for each polynomial order $d$. (a) Normalized absolute error ($\mathrm{NAE}^{(d)}$), showing the relative deviation from the unperturbed IPC as a function of the repetition count. The noise strength is fixed at $\sigma = 5\times 10^{-4}$, while the time series length varies with the number of repetitions such that the total number of data points is fixed at $10^6$. (b) Absolute error ($\mathrm{AE}^{(d)}$), showing the deviation from the unperturbed IPC under the same conditions. The results show that even at optimized repetition counts, the CROP method produces more accurate IPC estimates than the conventional averaging approach. }
  \label{fig:repeat_AE}
\end{figure}
Figures~\ref{fig:repeat_AE}(a) and (b) show $\mathrm{NAE}^{(d)}$ and $\mathrm{AE}^{(d)}$ for the IPC computed via repetition-based averaging, as functions of the repetition count. 
The total number of data points is fixed at $10^6$, so the time series length varies accordingly with the number of repetitions. 
These results show that the repetition count that minimizes the error can vary with a polynomial order $d$, reflecting a trade-off between the improved signal-to-noise ratio and the reduced time series length, which depends on the underlying unperturbed IPC distribution. 
For reference, $\mathrm{NAE}^{(d)}$ and $\mathrm{AE}^{(d)}$ obtained from CROP method at the same noise strength $\sigma = 5\times 10^{-4}$ are indicated by dotted lines; each color corresponds to a different polynomial order $d$. 
For each $d$, even when the repetition count is optimally chosen, the CROP method achieves lower $\mathrm{NAE}^{(d)}$ and $\mathrm{AE}^{(d)}$. 
This indicates that the CROP suppresses the effect of noise more efficiently than naive time-series averaging. 

\section{Application to quantum reservoirs}
\label{sec:quantum}
\subsection{System definition}
In the previous section, we demonstrated that the estimated noise-free IPC accurately approximates the unperturbed IPC of a classical discrete-time reservoir with quadratic nonlinearity. 
The CROP method is a natural fit for areas where noise is inherent or hard to avoid. 
Therefore, as an important application, we will now consider a quantum reservoir computing system. 
Quantum reservoir computing is a framework designed to achieve high performance by exploiting the quantum nature of physical systems. 
It benefits from a Hilbert space that scales exponentially with the number of nodes (i.e., qubits), as well as quantum correlations and entanglement that enrich the complexity of the underlying dynamics and information processing. 
However, a central tension for quantum reservoir computing is in the inherent difficulty of random state collapse under measurement, i.e. the sampling noise induced by the measurement problem of quantum mechanics.
The conventional IPC deviates significantly from the unperturbed IPC for a QRC, or a large amount of measurements is required to reconstruct it \cite{garcia2023scalable,palacios2024role,hu2023tackling,vcindrak2026memory,hahto2025smarter,liu2026practical}. 
This is especially punitive for current-generation experimental systems, where some sources of noise, like incoherent and readout errors, are limiting factors \cite{kubota2023temporal,yasuda2023quantum}.
We expect that the estimated noise-free IPC can significantly alleviate this problem, thereby helping to accurately identify promising physical substrates and configurations for quantum reservoir computing. 

As a representative example, we focus on a transverse-field Ising spin network system~\cite{Fujii2017}. 
Here, the dynamics of $N_q$ qubits is governed by the Hamiltonian, 
\begin{align}
    H = \sum_{i>j = 1}^{N_q} J_{ij}X_iX_j + \sum_{i=1}^{N_q} h_{\mathrm{ext}} Z_i, 
\end{align}
where $X_i, Z_i\in\mathbb{C}^{2^{N_q}\times2^{N_q}}$ are the Pauli operators for the $i$-th qubit, $J_{ij}$ represents the coupling strength between qubits $i$ and $j$, and $h_{\mathrm{ext}}$ denotes the strength of the external transverse field. 
At the beginning of each time step $k$, the first qubit is overwritten by the input-dependent density matrix $\rho_1 = |\psi_k\rangle\langle\psi_k|\in\mathbb{C}^2$, where the state vector is defined as 
\begin{align}
    |\psi_k\rangle = \sqrt{\frac{1-u[k]}{2}}|0\rangle + \sqrt{\frac{1+u[k]}{2}}|1\rangle. 
\end{align}
Here, the input $u[k]$ is drawn from i.i.d. uniform distribution over $[-1, 1]$, ensuring that the state vector is well-defined and standardly normalized. 
The total density matrix of the reservoir is given by 
\begin{align}
    \rho(k\Delta t) = e^{-iH\Delta t}\left(\rho_1 \otimes\mathrm{Tr}_1[\rho((k-1)\Delta t)]\right)e^{iH\Delta t},
\end{align}
where $\mathrm{Tr}_1[\cdot]$ denotes the partial trace over the first qubit, and $\Delta t$ is the duration of the subsequent unitary evolution. 

Here, we consider a system with $N_q = 6$ qubits. 
To construct the reservoir output, we select a specific set of $N = 63$ observables out of the $4^6-1 = 4095$ maximum independent degrees of freedom. 
This set of 63 observables consists of $3\times 6 = 18$ single-qubit operators ($X_i$, $Y_i$, and $Z_i$) and $3\times \tbinom{6}{2} = 45$ two-qubit correlation operators ($X_iX_j$, $Y_iY_j$, and $Z_iZ_j$). 
In numerical simulations, the ideal expectation value of each observable $A_n$ $(n = 1, \ldots, N)$ at time step $k$ is calculated by taking the trace with the density matrix, $\mathrm{Tr}[A_n\rho(k\Delta t)]$. 
In actual experiments, however, a single projective measurement on a basis (a `shot') yields only a binary measurement outcome (e.g., $+1$ or $-1$). 
Therefore, the same state must be repeatedly prepared and measured to estimate the expectation value. 

\subsection{Estimation of the noise-free IPC from noise-contaminated time series}
Here, we evaluate the effectiveness of our proposed CROP method to estimate the noise-free IPC in the quantum setup. 
We first compute the unperturbed IPC under an ideal condition as a benchmark. 
We then introduce sampling noise to the time series to evaluate how the IPC is degraded, and demonstrate how accurately the estimated noise-free IPC approximates the unperturbed IPC using the same noise-contaminated time series. 
Here, we consider the shot noise that corrupts the measurement outcomes as the primary source of noise. 
In our numerical simulations, this is implemented by simply adding Gaussian noise to the generated time series data. 
In an actual experiment, however, Gaussian noise is not directly added; instead, each single shot yields a binary measurement outcome that randomly flips. 
When averaged over a large number of shots, $S$, the sample mean approaches a Gaussian distribution with a scale of $1/\sqrt{S}$, justifying our simulation setup. 

\begin{figure}[htpb]
  \centering
  \includegraphics[width=0.43\textwidth]{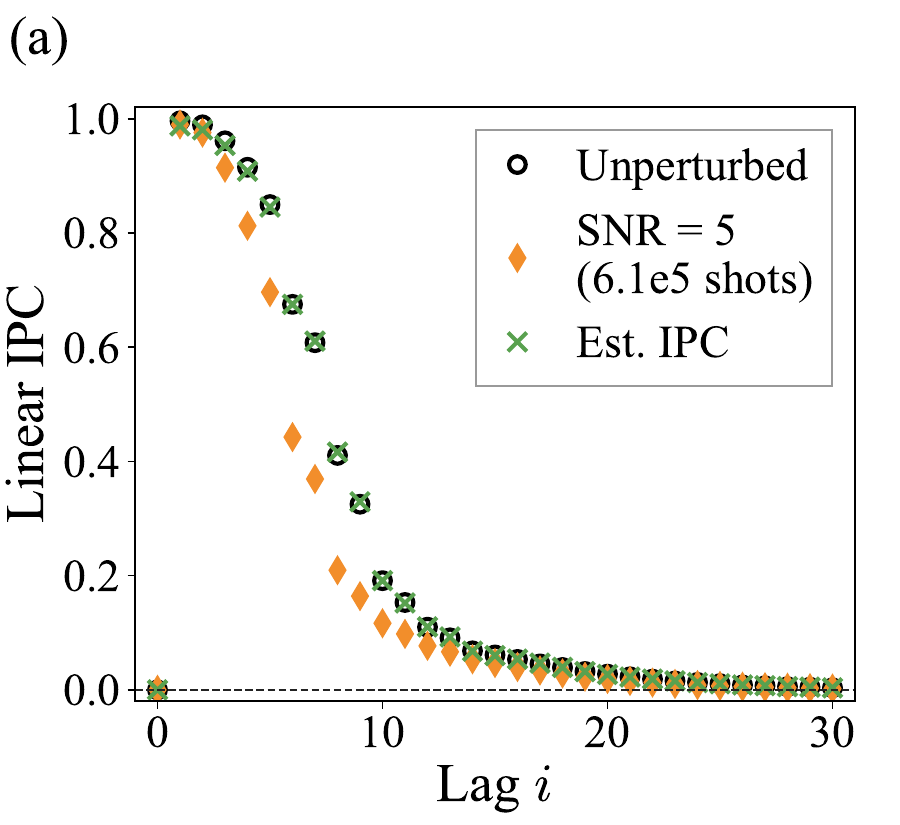}
  \includegraphics[width=0.43\textwidth]{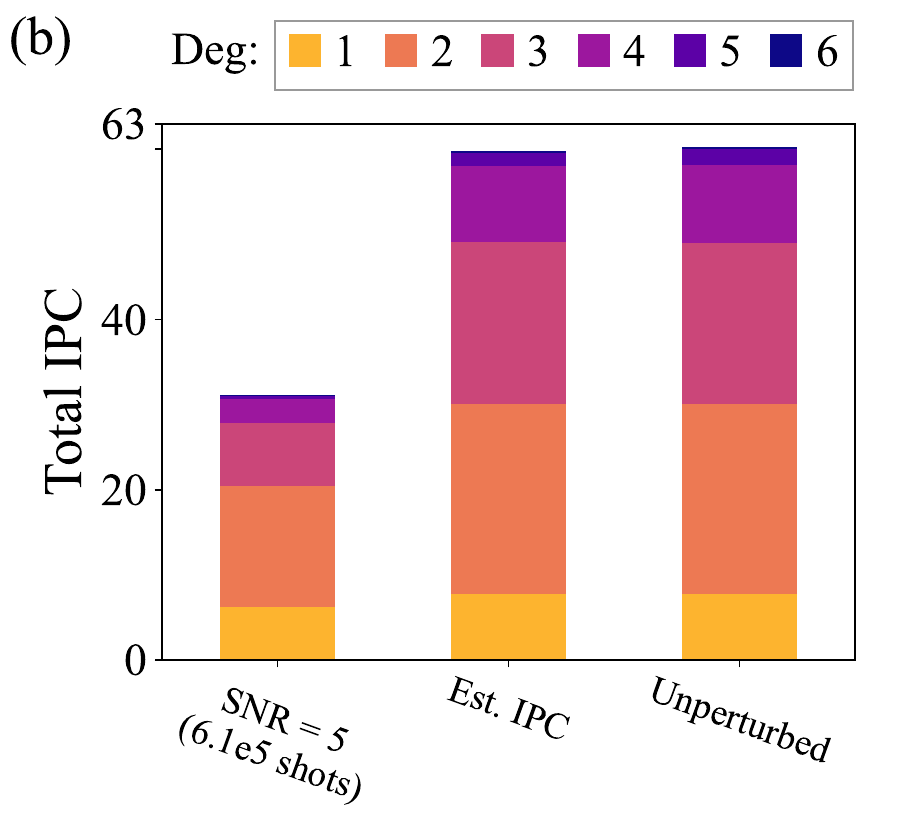}
  \caption{The IPC and the estimated noise-free IPC (`Est. IPC') of a 6-qubit quantum reservoir system with 63 observables under output noise. (a) The linear IPC $u[k-i]$: black circles, the unperturbed IPC; orange diamonds, the IPC directly computed from noisy time series of length $10^6$ with an SNR of $5$ (corresponding to approximately $6.1\times 10^5$ measurement shots); green crosses, the noise-free IPC estimated by the proposed method. (b) The total IPC up to the sixth degree from the same noisy data as in (a), with bars (left to right) showing the noise-degraded IPC, the estimated noise-free IPC, and the unperturbed IPC via CROP. }
  \label{fig:qrc_snr10}
\end{figure}
Figure~\ref{fig:qrc_snr10} shows the IPC and the estimated noise-free IPC calculated using $10^6$ data points. 
In Fig.~\ref{fig:qrc_snr10}(a), the black circles represent the unperturbed IPC for linear functions, and the orange diamonds represent the noise-degraded IPC. 
Here, the standard deviation of the unperturbed time series, calculated collectively across all observables, was $4.0\times 10^{-2}$. 
To achieve a signal-to-noise ratio (SNR) of $5$ the standard deviation of the added noise was set to $8.0\times 10^{-3}$, which corresponds to averaging approximately $6.1\times 10^5$ shots in hardware experiments. 
Comparing these results, we observe that the IPC significantly degrades across the entire range (more so than for the classical example shown in Fig.~\ref{fig:quad_5e-4}), including the short-lag regions of the target $u[k-i]$. 
In contrast, the green crosses represent the noise-free IPC estimated from the same noise-contaminated time series using CROP, closely matching the unperturbed baseline. 
Figure~\ref{fig:qrc_snr10}(b) shows the total IPC for each polynomial order up to the sixth, comparing (from left to right) the noise-degraded IPC, the estimated noise-free IPC via CROP, and the unperturbed ground-truth reference. 
For the unperturbed case, the total IPC reaches approximately $60$, which is close to the theoretical upper bound, $N = 63$. 
Unlike the example of classical reservoirs shown in Fig.~\ref{fig:quad_5e-4}(b), a slight deficit remains from the upper bound. 
This is because, as shown in Fig~\ref{fig:qrc_snr10}(a), the IPC decays slowly with increasing lag, leaving numerous small-IPC components that are easily underestimated due to overfitting. 
Compared to this unperturbed baseline, the noise-degraded IPC decreases across all orders, with its total dropping to approximately $31$. 
In contrast, the estimated noise-free IPC via CROP reaches a total of around $60$, exhibiting a nearly identical distribution particularly in the lower-order terms. 
This demonstrates that the CROP works effectively even in this quantum reservoir setup. 

\subsection{Comparison of the noise-free IPC estimation with shot-increased direct IPC evaluation}
In the quantum reservoir setup, we further investigate the inherent trade-off between the number of shots and the available time series length under a fixed total data resource. 
Unlike the classical case where ensemble averaging is an optional alternative, quantum reservoir computing fundamentally requires averaging over multiple shots at each time step to estimate continuous observables. 
For a fair comparison with our CROP, the total budget of data points is fixed at $6.1\times 10^{11}$, which is the product of the data length ($10^6$) and the baseline shots ($6.1\times 10^5$). 
This number of shots corresponds to the signal-to-noise ratio of $5$ in the setup shown in the previous section, reflecting the scale of hardware experiments. 
Consequently, increasing the number of shots suppresses the shot noise but simultaneously shortens the time series length, leading to a critical trade-off for accurate IPC evaluation. 

\begin{figure}[htbp]
  \centering
  \includegraphics[width=0.45\textwidth]{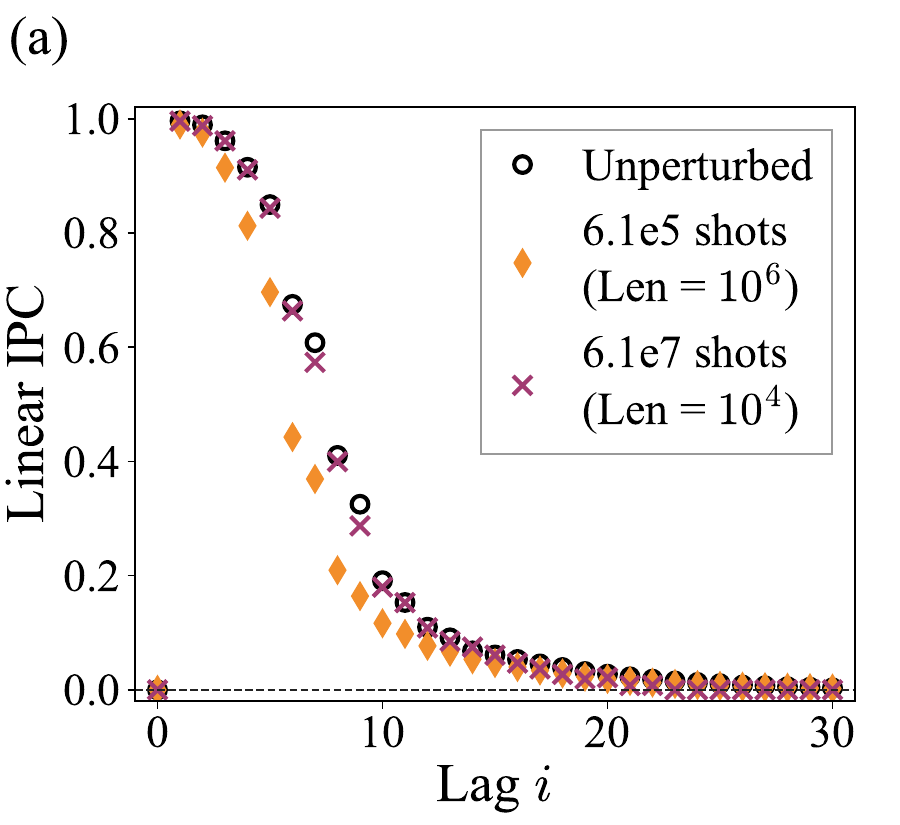}
  \includegraphics[width=0.45\textwidth]{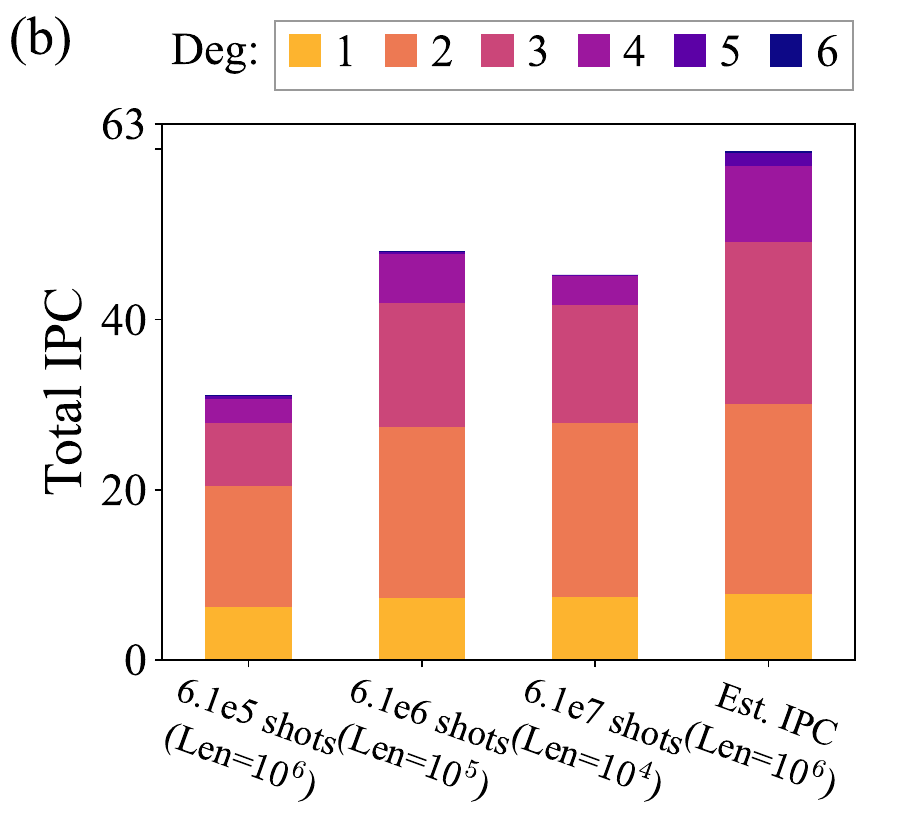}
  \caption{The IPC under different shot-count for a 6-qubit quantum reservoir system with 63 observables under output noise, with the total number of data points fixed. (a) The linear IPC $u[k-i]$: black circles, the unperturbed IPC; orange diamonds, the IPC computed from a reservoir time series of length $10^6$ obtained with $6.1\times10^5$ shots per time step; purple crosses, that for a series of length $10^4$ with $6.1\times 10^7$ shots. (b) The total IPC up to the sixth degree, with bars (left to right) showing the IPC for $6.1\times 10^5$, $6.1\times 10^6$, $6.1\times 10^7$ shots per time step (series length of $10^6$, $10^5$, and $10^4$), and our estimated noise-free IPC for reference.}
  \label{fig:qrc_repeat}
\end{figure}
Figure~\ref{fig:qrc_repeat} shows the IPC evaluated under various shot-count allocations while keeping the total number of data points fixed. 
In Fig.~\ref{fig:qrc_repeat}(a), the orange diamonds represent the linear IPC computed with $6.1\times 10^5$ shots per time step (series length of $10^6$) at an signal-to-noise ratio of $5$, which corresponds to the setup shown in Fig.~\ref{fig:qrc_snr10}(a). 
The purple crosses show the result when the number of shots is increased to $6.1\times 10^7$, whereas the series length is shortened to $10^4$ to maintain the total data budget. 
The black circles provide the unperturbed IPC for reference. 
As shown in the plot, the purple crosses are significantly closer to the unperturbed IPC compared to the orange diamonds. 
This indicates that increasing the number of shots suppresses the output noise, thereby bringing the evaluated IPC much closer to the ideal unperturbed baseline. 
However, as shown in Fig.~\ref{fig:qrc_repeat}(b), simply increasing the number of shots does not indefinitely improve the results.
When the number of shots is increased from $6.1\times 10^5$ to $6.1\times 10^6$, the total IPC improves to approximately $48$. 
Conversely, further increasing the shots to $6.1\times 10^7$ leads to a degradation in the total IPC. 
This degradation is attributed to the excessively shortened time series, which impairs evaluation accuracy and outweighs the benefit of noise reduction, mirroring the same effect already observed for the classical system in Fig.~\ref{fig:quad_repeat}.
In contrast, the estimated noise-free IPC via CROP calculated using a series length of $10^6$, shown for reference, is closer to the unperturbed IPC than any of these directly evaluated results, showing that CROP gives more accurate estimates than averaging-based noise reduction for the quantum system as well.

\section{Conclusion}
\label{sec:conclusion}

We have demonstrated how noise affects the information processing capacity of input-driven dynamical systems and how the noise-free IPC can be reconstructed from noisy observations.

First, we distinguished the unperturbed dynamics $r$ from the noise-free component $\bar{r}$ of the stochastic dynamics, i.e., the orthogonal projection of the observed state $r'$ onto the subspace of functions depending only on the input history. 
The two do not, in general, coincide: the projection removes the fluctuation $\xi$ orthogonal to this subspace rather than reproducing the unperturbed dynamics. In a minimal one-dimensional reservoir, additive input noise redistributes capacity from the third- to the first-order Legendre polynomial, so that $\bar{r}$ acquires a target function absent in $r$ while the total IPC of $\bar{r}$ is conserved. 
The IPC of the noisy state factorizes into this redistributed capacity and a term measuring the decay of the signal strength. 
Consequently, noise can enhance the performance on a particular task even as the total capacity decreases. 

Second, we introduced covariance reconstruction by orthogonal projection (CROP). 
Because $\xi$ is orthogonal to every noise-independent function, the cross-correlations of $r'$ with a complete orthonormal basis of target functions extract only the component lying in $\mathcal{M}_u$, and their outer products reconstruct the covariance of $\bar{r}$ from noisy data. 
This yields $C(z\mid\bar{r})$ without requiring a detailed model of either the system or the noise. 
The estimate is exact when $\bar{r}$ lies in the span of the truncated basis.

Third, we validated CROP numerically, on both a classical 100-node echo state network with additive process noise and 6-qubit transverse-field Ising reservoir under shot noise. 
CROP can successfully reconstruct most of the IPC even when the direct measurement would yield only a greatly reduced capacity.
At fixed total measurement budget, CROP is more accurate than repetition-based averaging: in the classical case for every polynomial order even at the optimal repetition count, and in the quantum case for every tested shot allocation. 
Averaging faces an intrinsic trade-off as more repetitions improve the signal-to-noise ratio but shorten the time series available for the IPC evaluation. 
CROP uses the full time series and removes the noise contribution in the covariance instead.

Several limitations remain. CROP estimates the IPC of $\bar{r}$ at a fixed noise strength and distribution, not that of $r$; predicting the redistributory effect without a model of the system is beyond the scope of this work. 
The reconstruction depends on the truncation $L$ of the basis, for which we have no general error bound: too small an $L$ underestimates the covariance, while increasing $L$ too much also degrades the estimate. 
However, since $\bar{r}$ is precisely what ensemble averaging would yield in the limit of infinitely many repetitions, CROP provides from a single noisy time series what averaging can only approach at the cost of data length. 

Using the CROP method to estimate the information processing capability of noisy physical systems has the potential to shape the development of physical computing hardware. 
For physical platforms where noise is fundamentally unavoidable, in particular quantum reservoir computing, the estimated noise-free IPC serves as a systematic and sample-efficient benchmark to identify promising physical substrates, network architectures, or parameter regimes from finite observations.

\begin{acknowledgments}
This study was supported in part by a Grant-in-Aid for Transformative Research Areas (A) (JP22H05197), Grant-in-Aid for Scientific Research(S) (JP26K21733) and a Grant-in-Aid for JSPS Fellows (JP25KJ0758).
JN acknowledges The Finnish Quantum Institute (InstituteQ).
LJ acknowledges funding from the Nexus program of the Carl-Zeiss-Stiftung.
\end{acknowledgments}

\appendix

\section{Derivation of the IPC for the one-dimensional reservoir \label{sec:r=barr_calc}}
In this section, we provide the detailed calculation of the IPC for the one-dimensional reservoir introduced in Sec.~\ref{sec:r=barr}. 
For notational simplicity, throughout this section, we use $u$ and $v$ to denote $u[k]$ and $v[k]$, respectively. 
The input $u$ is assumed to be uniformly distributed on $[-1, 1]$. 
The observed reservoir state is defined as 
\begin{align}
    r'(u, v) = r(u+v), 
\end{align}
where 
\begin{align}
    r(u) = \frac{1}{2}(5u^3 - 3u) = \frac{1}{\sqrt{7}}P_3(u). 
\end{align}
Let $v$ be a zero-mean Gaussian random variable with standard deviation $\sigma$. 
Expanding $r'(u,v) = r(u+v)$ gives
\begin{align}
    r'(u,v) = r(u) + \frac{15}{2}u^2v + \frac{15}{2}uv^2 + \frac{5}{2}v^3 - \frac{3}{2}v. 
\end{align}
The IPC of a one-dimensional reservoir is written as 
\begin{align}
    C(z\mid r') = \frac{(\langle zr'\rangle_{u,v})^2}{\langle z^2\rangle_u\langle r'^2\rangle_{u,v}}. 
\end{align}
We evaluate each term in the above expression for the target functions $P_1(u) = \sqrt{3}u$ and $P_3(u)$. 
Expanding $r'^2$, and noting that $\langle v\rangle_v = \langle v^3\rangle_v = \langle v^5\rangle_v = 0$, we obtain
\begin{align}
    &\langle r'^2\rangle_{u,v} \notag\\
    &= \langle r^2\rangle_u +\frac{225}{4}\langle u^4\rangle_u\langle v^2\rangle_v
+\frac{225}{4}\langle u^2\rangle_u\langle v^4\rangle_v \notag\\
&\quad +\frac{25}{4}\langle v^6\rangle_v
+\frac{9}{4}\langle v^2\rangle_v + 15\langle ru\rangle_u\langle v^2\rangle_v\notag\\
&\quad + \frac{75}{2}\langle u^2\rangle_u\langle v^4\rangle_v- \frac{45}{2}\langle u^2\rangle_u\langle v^2\rangle_v -\frac{15}{2}\langle v^4\rangle_v.
\end{align}
Here, 
\begin{align}
    \langle r^2\rangle_u = \frac{1}{7},\quad \langle u^2\rangle_u = \frac{1}{3}, \quad\langle u^4\rangle_u = \frac{1}{5}, 
\end{align}
and 
\begin{align}
    \langle ru\rangle_u = \frac{1}{2} (5\langle u^4\rangle_u - 3\langle u^2\rangle_u) = 0. 
\end{align}
For Gaussian noise, 
\begin{align}
    \langle v^2\rangle_v = \sigma^2, \quad \langle v^4\rangle_v = 3\sigma^4, \quad \langle v^6\rangle_v = 15\sigma^6, 
\end{align}
and hence 
\begin{align}
    \langle r'^2\rangle_{u,v} = \frac{1}{7} + 6\sigma^2 + \frac{285}{4}\sigma^4 + \frac{375}{4}\sigma^6. 
\end{align}
For the first-order target $P_1$, 
\begin{align}
    \langle P_1 r'\rangle_{u,v} = \frac{15}{2}\langle P_1u\rangle_u\langle v^2\rangle_v = \frac{5\sqrt{3}}{2}\sigma^2. 
\end{align}
Therefore, 
\begin{align}
    C(P_1(u)\mid r') &= \frac{\frac{75}{4}\sigma^4}{\frac{1}{7} + 6\sigma^2 + \frac{285}{4}\sigma^4 + \frac{375}{4}\sigma^6 }.
\end{align}
For the third-order target $P_3$, 
\begin{align}
    \langle P_3r'\rangle_{u,v} = \langle P_3r\rangle_u = \frac{1}{\sqrt{7}}. 
\end{align}
Hence, 
\begin{align}
    C(P_3(u)\mid r') &= \frac{\frac{1}{7}}{\frac{1}{7} + 6\sigma^2 + \frac{285}{4}\sigma^4 + \frac{375}{4}\sigma^6 }. 
\end{align}

\section{Analytical and recursive solution of a quadratic system\label{sec:analytical}}
For the classical reservoir with quadratic nonlinearity presented in Sec.~\ref{subsec:reconst}, we calculate the unperturbed IPC as a baseline for comparison with the noise-degraded IPC and the estimated noise-free IPC via CROP. 
Here, we can use the unperturbed dynamics as a proxy here, as we heuristically observed that for this system, the difference between unperturbed dynamics and the noise-free part $\tilde{r}$ for any fixed noise strength is small compared to the components entirely missed due to finite-size data length in the reconstruction methods.

As a preparation for this IPC evaluation, we analytically derive the Wiener-Volterra expansion of the discrete-time dynamical system, 
\begin{align}
    r[k] = Ar[k-1] - (r[k-1]\odot r[k-1]) + Bu[k], 
\end{align}
where $r[k]\in\R^N$ denotes the system state at time $k$, $A\in\R^{N\times N}$ is the state transition matrix, $B\in\R^N$ is the input vector, and $u[k]\in\mathcal{U}\subseteq\R$ is a scalar input. 
The symbol $\odot$ denotes the Hadamard (element-wise) product. 

We decompose the state into contributions of different input orders, as 
\begin{align}
    r[k] = \sum_{d=1}^\infty \Psi_d\left(u^{(\infty)}[k]\right), 
\end{align}
where $u^{(\infty)}[k] = (u[k], u[k-1], \ldots)\in\mathcal{U}^{\Z}$ represents the entire input history up to time $k$, and $\Psi_d(\cdot): \mathcal{U}^{\Z}\to\R^N$ denotes the $d$-th degree functional with respect to the input. 
In the following, 
\begin{align}
    \Psi_d[k] := \Psi_d\left(u^{(\infty)}[k]\right)
\end{align}
for simplicity. 

We first consider the first-order component. 
Since the quadratic term does not contribute at this order, $\Psi_1[k]$ satisfies 
\begin{align}
    \Psi_1[k] = A\Psi_1[k-1] + Bu[k]. 
\end{align}
Solving this recursion yields 
\begin{align}
    \Psi_1[k] = A^k\Psi_1[0] + \sum_{i=0}^{k-1}A^iBu[k-i]. 
\end{align}
Assuming that the system is stable, i.e., the spectral radius of $A$ satisfies $\rho(A) < 1$, the influence of the initial condition vanishes as $k\to\infty$. 
In this case, 
\begin{align}
    \Psi_1[k] = \sum_{i=0}^{\infty}A^iBu[k-i]. 
\end{align}

We next consider the second-order component. 
From the quadratic interaction term, $\Psi_2[k]$ satisfies 
\begin{align}
    \Psi_2[k] = A\Psi_2[k-1] -\Psi_1[k-1]\odot \Psi_1[k-1]. 
\end{align}
Solving this recursion and taking the limit $k\to\infty$, the influence of the initial condition vanishes under the assumption $\rho(A) < 1$, yielding
\begin{align}
    \Psi_2[k] = - \sum_{i=0}^{\infty}A^i(\Psi_1[k-1-i]\odot \Psi_1[k-1-i]).
\end{align}
Substituting the expression for $\Psi_1[k]$ derived above, we rewrite the term $\Psi_1[k-1-i]$ as
\begin{align}
    \Psi_1[k-1-i]&=\sum_{j=0}^{\infty}A^jBu[k-1-i-j]\notag\\
    &= \sum_{s=1+i}^{\infty}A^{s-1-i}Bu[k - s]\notag\\
    &= \sum_{s=1+i}^{\infty}g_{s-1-i}u[k-s],
\end{align}
where we have introduced the change of variable $s := 1 + i + j$ and defined $g_j := A^jB$ for convenience. 
Substituting this into the expression for $\Psi_2[k]$, we obtain
\begin{align}
    &\Psi_2[k] \notag\\
    &= -\sum_{i=0}^{\infty}A^i\sum_{s,t=1+i}^{\infty}(g_{s-1-i}\odot g_{t-1-i})u[k-s]u[k-t]
\end{align}
By rearranging the order of summation, the coefficient of $u[k-s]u[k-t]$, denoted by $c_{st}$ is given by 
\begin{align}
    c_{st} = - \sum_{i=0}^{\min(s,t)-1}A^i(g_{s-1-i}\odot g_{t-1-i}).
\end{align}

We next consider the general $d$-th order component with $d\ge 2$. 
Due to the quadratic nonlinearity, the $d$-th order component is generated by interactions between lower-order components whose orders sum to $d$. 
Thus $\Psi_d[k]$ satisfies the following recursion: 
\begin{align}
    \Psi_d[k] = A\Psi_d[k-1] - \sum_{l+m = d}\Psi_l[k-1]\odot \Psi_m[k-1].
\end{align}
Expanding it recursively and taking the limit $k\to\infty$, the influence of the initial condition vanishes under the assumption $\rho(A) < 1$, yielding 
\begin{align}
    \Psi_d[k] = -\sum_{i=0}^{\infty}A^i\sum_{l+m=d}\Psi_l[k-1-i]\odot \Psi_m[k-1-i]. 
\end{align}
Thus, higher-order components can be computed sequentially through the recursive structure of the system. 
In practice, the expansion is truncated at a finite delay index and polynomial order, and all terms within the range are obtained recursively. 

\section{Wiener-Volterra time series and the information processing capacity\label{sec:volterra}}
Following the approach demonstrated in the previous section, we assume that the Wiener-Volterra expansion of the input-driven reservoir with respect to the input has been obtained. 
The resulting expansion can be expressed as 
\begin{align}
    r[k] = \sum_{\nu}c_{\nu}\phi_{\nu}\left(u^{(\infty)}[k]\right),
\end{align}
where $\nu$ is a multi-index specifying the combination of delay indices and polynomial orders, $\phi_{\nu}$ denotes the corresponding monomial of the input history, and $c_{\nu}\in\R^{N}$ are the expansion coefficient vectors. 
Based on this expansion, we compute the IPC for a given target function $z$. 
This approach is mathematically analogous to that in~\cite{Kubota2021}.
In that work, the expansion is referred to as polynomial chaos; however, to avoid confusion with the notion of chaos in reservoir computing, where the IPC was originally introduced, we adopt the term Wiener-Volterra expansion. 

To ensure consistency with the finite-data setting, we assume that the input series $u[k]$ is i.i.d. according to a specific probability distribution, and choose target functions from a basis orthonormal with respect to this distribution. 
For the uniform distribution on $[-1,1]$, we use normalized Legendre polynomials as the orthonormal basis. 
Since these basis functions span the polynomial space, each monomial $\phi_{\nu}$ can be expressed as a linear combination of them. 
Accordingly, the Wiener-Volterra expansion can be rewritten as 
\begin{align}
    r[k] = \sum_{\nu}\hat{c}_{\nu}\Phi_{\nu}\left(u^{(\infty)}[k]\right),
\end{align}
where $\Phi_{\nu}$ denotes products of normalized Legendre polynomials of delayed inputs, and $\hat{c}_{\nu}\in\R^N$ are the coefficients obtained by applying the corresponding linear transformation to $c_{\nu}$. 
The functions $\{\Phi_{\nu}\}$ form an orthonormal basis with respect to the assumed input distribution, i.e., 
\begin{align}
    \langle\Phi_{\nu}\Phi_{\mu}\rangle_{u} = \delta_{\nu\mu}, 
\end{align}
where $\delta_{\nu\mu}$ denotes the Kronecker delta. 

From the definition, the IPC can be expressed as in Eq.~\eqref{eq:czr}. 
We consider the target function $z = \Phi_{\nu}$ and first examine the special case of a one-dimensional reservoir for simplicity. 
Using 
\begin{align}
    r[k] = \sum_{\nu}\hat{c}_{\nu}\Phi_{\nu}, \quad\hat{c}_{\nu}\in\R, 
\end{align}
we obtain
\begin{align}
    \langle \Phi_{\nu} r\rangle_u &= \left\langle\Phi_{\nu}\sum_{\mu}\hat{c}_{\mu}\Phi_{\mu}\right\rangle_u\notag\\
    &=\sum_{\mu}\hat{c}_{\mu}\delta_{\nu\mu} =\hat{c}_{\nu}, 
\end{align}
\begin{align}
    \langle rr\rangle_u &= \left\langle\left(\sum_{\mu}\hat{c}_{\mu}\Phi_{\mu}\right)^2\right\rangle_u\notag\\
    &= \sum_{\nu,\mu}\hat{c}_{\nu}\hat{c}_{\mu}\delta_{\nu\mu} = \sum_{\mu}\hat{c}_{\mu}^2.
\end{align}
Therefore, 
\begin{align}
    C(\Phi_{\nu}\mid r) = \frac{\hat{c}_{\nu}^2}{\sum_{\mu}\hat{c}_{\mu}^2}. 
\end{align}
Thus, for a one-dimensional reservoir, the IPC associated with each basis function is proportional to the squared Wiener-Volterra coefficient. 

We now turn to the general multi-dimensional case. 
For a multi-dimensional reservoir, the expansion coefficients $\hat{c}_{\nu}\in\R^N$ are vectors. 
Using the orthonormality of the basis functions, 
\begin{align}
    \left\langle\Phi_{\nu}r\right\rangle_u &= \left\langle\Phi_{\nu}\sum_{\mu}\hat{c}_{\mu}\Phi_{\mu}\right\rangle_u= \hat{c}_{\nu},
\end{align}
\begin{align}
    \langle rr^{\mathsf{T}}\rangle_u &= \left\langle\left(\sum_{\nu}\hat{c}_{\nu}\Phi_{\nu}\right)\left(\sum_{\mu}\hat{c}_{\mu}\Phi_{\mu}\right)^{\mathsf{T}}\right\rangle_u\notag\\
    &= \sum_{\nu,\mu}\hat{c}_{\nu}\hat{c}_{\mu}^{\mathsf{T}}\delta_{\nu\mu} = \sum_{\mu}\hat{c}_{\mu}\hat{c}_{\mu}^{\mathsf{T}}. 
\end{align}
Therefore, 
\begin{align}
    C(\Phi_{\nu}\mid r) &= \hat{c}_{\nu}^{\mathsf{T}}\left(\sum_{\mu}\hat{c}_{\mu}\hat{c}_{\mu}^{\mathsf{T}}\right)^{\dagger}\hat{c}_{\nu}.
\end{align}
Thus, in the multi-dimensional case, the IPC remains quadratic in the Wiener-Volterra coefficients, but the scalar normalization in the one-dimensional case is replaced by the pseudo-inverse of the coefficient covariance matrix. 

\section{Finite-sample bias of the IPC and correction method\label{sec:samplebias}}
In this section, we discuss finite-sample effects in evaluation of the IPC. 
We review the finite-sample bias arising in covariance-based IPC evaluation, and a previously proposed bias-correction method~\cite{ramachandran2026information}. 
In addition, we show the corresponding bias in the train/test split formulation of the IPC. 

We begin with the covariance-based IPC formulation studied in~\cite{ramachandran2026information}, where both the optimal read-out weight and the IPC are evaluated using the same finite dataset.
For consistency with the setting in~\cite{ramachandran2026information}, and for simplicity, the following analysis is presented for an extreme learning machine. 
In this setting, the system state depends only on the current input and not on its history, so that the sequence $\{r[t]\}_{t=1}^T$ can be treated as i.i.d. samples. 
We further assume that these samples have finite second moments, which is required for the central limit theorem used in the subsequent analysis. 
Even for systems with temporal correlations, the arguments presented below are expected to extend under appropriate assumptions on the temporal correlations. 

As shown in Eq.~\eqref{eq:czr}, the IPC defined in terms of ensemble averages is obtained by substituting the optimal output weight  
\begin{align}
    w^* := (\langle rr^{\mathsf{T}}\rangle_u)^{-1}\langle zr\rangle_u 
\end{align}
into the definition of the IPC, yielding 
\begin{align}
    C(z\mid r) =\frac{\langle zr\rangle_u^{\mathsf{T}}(\langle rr^{\mathsf{T}}\rangle_u)^{-1}\langle zr\rangle_u}{\langle z^2\rangle_u}.  
\end{align}
Here, we assume that $\langle rr^{\mathsf{T}}\rangle_u$ is invertible for simplicity. 
For a target function $z$, we define the residual
\begin{align}
    \epsilon := z - w^{*\mathsf{T}}r. 
\end{align}
By construction, $\langle \epsilon r\rangle_u = 0$. 
The corresponding finite-sample IPC is defined as
\begin{align}
    C_T(z\mid r) := \frac{\langle zr\rangle_T^{\mathsf{T}}(\langle rr^{\mathsf{T}}\rangle_T)^{-1}\langle zr\rangle_T }{\langle z^2\rangle_u}, 
\end{align}
where $\langle\cdot\rangle_T$ denotes the empirical time average over a finite dataset of length $T$. 
For simplicity, we assume that the target variance $\langle z^2\rangle_u$ is known analytically, as is typically the case for commonly used basis functions such as Legendre polynomials. 

To quantify the finite-sample bias, we analyze the deviation of this $C_T(z\mid r)$ from its infinite-data counterpart $C(z\mid r)$ as a function of the dataset size $T$. 
Using $z = w^{*\mathsf{T}}r + \epsilon$, the empirical cross-correlation can be decomposed as $\langle zr\rangle_T = \langle rr^{\mathsf{T}}\rangle_Tw^* + \langle\epsilon r\rangle_T$. 
We introduce the orthogonalized state 
\begin{align}
    \hat{r} := (\langle rr^{\mathsf{T}}\rangle_u)^{-1/2} r,
\end{align}
and define the fluctuation terms 
\begin{align}
    a_T := \langle \hat{r}\hat{r}^{\mathsf{T}}\rangle_T - I, \quad b_T := \langle \epsilon\hat{r}\rangle_T. 
\end{align}
Then, the finite-sample IPC can be rewritten as
\begin{align}
    &C_T(z\mid r)= \notag\\
    & \frac{\hat{w}^{*\mathsf{T}}(I + a_T)\hat{w}^* + 2\hat{w}^{*\mathsf{T}}b_T + b_T^{\mathsf{T}}(I + a_T)^{-1}b_T}{\langle z^2\rangle_u},
\end{align}
where $\hat{w}^* := (\langle rr^{\mathsf{T}}\rangle_u)^{1/2}w^*$. 

We first characterize the scaling of the fluctuation terms $a_T$ and $b_T$. 
Consider an arbitrary entry of $a_T$, 
\begin{align}
    (a_T)_{ij} = \frac{1}{T} \sum_{t=1}^T (\hat{r}_i[t]\hat{r}_j[t] - \delta_{ij}). 
\end{align}
Since 
\begin{align}
    \langle \hat{r}_i\hat{r}_j - \delta_{ij}\rangle_u = 0, 
\end{align}
each summand has zero mean. 
As $\{r[t]\}_t$ are i.i.d, the sequence $\hat{r}_i[t]\hat{r}_j[t] - \delta_{ij}$ 
Thus, the central limit theorem implies
\begin{align}
    \sqrt{T}(a_T)_{ij} \xrightarrow[]{d} \mathcal{N}\left(0, \mathrm{Var}(\hat{r}_i\hat{r}_j - \delta_{ij})\right), 
\end{align}
where $\xrightarrow[]{d}$ denotes convergence in distribution. 
Therefore, 
\begin{align}
    (a_T)_{ij} &= \frac{1}{\sqrt{T}} \left(\frac{1}{\sqrt{T}} \sum_{t=1}^T (\hat{r}_i[t]\hat{r}_j[t] -\delta_{ij})\right) \notag\\
    &= O_p(T^{-1/2}), 
\end{align}
where $X_T = O_p(f(T))$ denotes that $X_T/f(T)$ remains bounded in probability as $T\to\infty$. 
The same argument applies to each entry of $b_T$, yielding 
\begin{align}
    (b_T)_i = O_p(T^{-1/2}). 
\end{align}

Since $(a_T)_{ij} = O_p(T^{-1/2})$, we have $a_T^2 = O_p(T^{-1})$. 
Using the Neuman series expansion, 
\begin{align}
    (I + a_T)^{-1} = I - a_T + O_p(T^{-1}), 
\end{align}
and by substituting this, we obtain
\begin{align}
    &C_T(z\mid r) \notag \\
    &= \frac{\hat{w}^{*\mathsf{T}}a_T\hat{w}^* + 2\hat{w}^{*\mathsf{T}}b_T + b_T^{\mathsf{T}}b_T - b_T^{\mathsf{T}}a_Tb_T}{\langle z^2\rangle_u} \notag\\
    &\quad+ C(z\mid r) + O_p(T^{-2}). 
\end{align}
By construction, $\langle \hat{r}_i\hat{r}_j - \delta_{ij}\rangle_u = 0$ and $\langle \epsilon\hat{r}_i\rangle_u = 0$, so that $\langle (a_T)_{ij}\rangle_u = 0$ and $\langle (b_T)_i\rangle_u = 0$. 
Taking the expectation with respect to the input $u$, the linear fluctuation terms, $\hat{w}^{*\mathsf{T}}a_T\hat{w}^*$ and $2\hat{w}^{*\mathsf{T}}b_T$ vanish. 
Moreover, 
\begin{align}
    &\langle b_T^{\mathsf{T}}a_Tb_T\rangle_u\notag\\
    &= \frac{1}{T^3}\sum_{t,q,s=1}^T\langle\epsilon[t]\hat{r}[t]^{\mathsf{T}}(\hat{r}[q]\hat{r}[q]^{\mathsf{T}} - I)\epsilon[s]\hat{r}[s] \rangle_u, 
\end{align}
where $t$, $q$, and $s$ denote time indices. 
Again, from $\langle (a_T)_{ij}\rangle_u = 0$ and $\langle (b_T)_i\rangle_u = 0$ and by independence of samples, all terms vanish unless $t = q = s$. 
Therefore, 
\begin{align}
    \langle b_T^{\mathsf{T}} a_Tb_T\rangle_u = O(T^{-2}). 
\end{align}
Thus, up to the leading finite-sample bias, 
\begin{align}
    \langle C_T(z\mid r)\rangle_u = C(z\mid r) + \frac{\Delta(z\mid r)}{T\langle z^2\rangle_u} + O(T^{-2}),
\end{align}
where $\Delta(z\mid r) := \langle \epsilon^2 \hat{r}^{\mathsf{T}}\hat{r}\rangle_u\ge 0$. 
Thus, the covariance-based IPC has a positive finite-sample bias at the order of $1 / T$. 

In practical IPC evaluations, the covariance-based formulation is widely used. 
To mitigate the positive bias, it is common to apply a thresholding procedure, in which IPC values below a prescribed threshold are set to zero. 
This is primarily intended to suppress spurious nonzero IPC values that arise from target functions whose true IPC is zero. 
However, it does not correct the bias of target functions with nonzero IPC. 
To address this issue, \cite{ramachandran2026information} proposed a bias-corrected estimator based on Richardson extrapolation, 
\begin{align}
    \hat{C}_T := 2C_T - C_{T/2}, 
\end{align}
where $C_{T/2}$ denotes the IPC evaluated using half of the available samples. 
Since the leading finite-sample bias scales as $1/T$, this combination cancels the first-order bias term, yielding
\begin{align}
    \langle\hat{C}_T\rangle_u = C(z\mid r) + O(T^{-2}). 
\end{align}

Next, we discuss the finite-sample bias in the train/test split formulation of the IPC. 
Let the training dataset and test dataset consist of $T_{\mathrm{tr}}$ and $T_{\mathrm{te}}$ samples, respectively. 
Furthermore, let $\langle\cdot\rangle_{\mathrm{tr}}$ and $\langle\cdot\rangle_{\mathrm{te}}$ denote empirical averages over the training and test datasets. 
The read-out weight is estimated from the training dataset as $w_{\mathrm{tr}}^* := (\langle rr^{\mathsf{T}}\rangle_{\mathrm{tr}})^{-1}\langle zr\rangle_{\mathrm{tr}}$. 
The resulting IPC is defined by 
\begin{align}
    C_{\mathrm{te}}(z\mid r) := 1 - \frac{\langle (w_{\mathrm{tr}}^{*\mathsf{T}}r - z)^2\rangle_{\mathrm{te}}}{\langle z^2\rangle_u}. 
\end{align}
To analyze the finite-sample bias, we define
\begin{align}
    \delta w := w_{\mathrm{tr}}^* - w^*, 
\end{align}
where $w^* = (\langle rr^{\mathsf{T}}\rangle_u)^{-1}\langle zr\rangle_u$ is the optimal output weight. 
Using $z = w^{*\mathsf{T}}r + \epsilon$, the prediction error on the test dataset becomes $w_{\mathrm{tr}}^{\mathsf{T}}r -z = \delta w^{\mathsf{T}}r - \epsilon$. 
Therefore, 
\begin{align}
    &C_{\mathrm{te}}(z\mid r) \notag\\
    &= 1 - \frac{\langle\epsilon^2\rangle_{\mathrm{te}}}{\langle z^2\rangle_u} + \frac{2\delta w^{\mathsf{T}}\langle\epsilon r\rangle_{\mathrm{te}}}{\langle z^2\rangle_u} - \frac{\delta w^{\mathsf{T}}\langle rr^{\mathsf{T}}\rangle_{\mathrm{te}}\delta w}{\langle z^2\rangle_u}. 
\end{align}
We now take the expectation of the above with regard to the input $u$. 
Since the test samples are independently drawn from the underlying distribution, 
\begin{align}
    1 - \frac{\langle\langle\epsilon^2\rangle_{\mathrm{te}}\rangle_u}{\langle z^2\rangle_u} = 1 - \frac{\langle \epsilon^2\rangle_u}{\langle z^2\rangle_u} = C(z\mid r). 
\end{align}
Regarding the cross term, 
\begin{align}
    \delta w = w_{\mathrm{tr}}^* - w^* = (\langle rr^{\mathsf{T}}\rangle_{\mathrm{tr}})^{-1}\langle \epsilon r\rangle_{\mathrm{tr}}.
\end{align}
Since the training and test datasets are independent, 
\begin{align}
    \langle\delta w^{\mathsf{T}}\langle\epsilon r\rangle_{\mathrm{te}}\rangle_u = \langle\delta w\rangle_u^{\mathsf{T}}\langle\langle\epsilon r\rangle_{\mathrm{te}}\rangle_u = 0. 
\end{align}
Thus, 
\begin{align}
    \langle C_{\mathrm{te}}(z\mid r)\rangle_u = C(z\mid r) - \frac{\langle \delta w^{\mathsf{T}}\langle rr^{\mathsf{T}}\rangle_{\mathrm{te}}\delta w\rangle_u}{\langle z^2\rangle_u}.
\end{align}
We next evaluate the quadratic term in the weight error. 
Define the whitened state $\hat{r} = \langle rr^{\mathsf{T}}\rangle_u^{-1/2}r$, and 
\begin{align}
    a_{\mathrm{tr}} &:= \langle \hat{r}\hat{r}^{\mathsf{T}}\rangle_{\mathrm{tr}} - I, \quad b_{\mathrm{tr}} := \langle\epsilon\hat{r}\rangle_{\mathrm{tr}},\\
    a_{\mathrm{te}} &:= \langle \hat{r}\hat{r}^{\mathsf{T}}\rangle_{\mathrm{te}} - I.
\end{align}
Then, we obtain 
\begin{align}
    \delta w = (\langle rr^{\mathsf{T}}\rangle_u)^{-1/2}(I + a_{\mathrm{tr}})^{-1}b_{\mathrm{tr}}.
\end{align}
Substituting this into the quadratic term gives 
\begin{align}
     &\delta w^{\mathsf{T}}\langle rr^{\mathsf{T}}\rangle_{\mathrm{te}}\delta w\notag\\
     &= b_{\mathrm{tr}}^{\mathsf{T}}(I + a_{\mathrm{tr}})^{-1}(I + a_{\mathrm{te}})(I + a_{\mathrm{tr}})^{-1}b_{\mathrm{tr}}.
\end{align}
Using the Neumann series expansion, 
\begin{align}
    (I + a_{\mathrm{tr}})^{-1} = I - a_{\mathrm{tr}} + O_p(T_{\mathrm{tr}}^{-1}), 
\end{align}
we obtain
\begin{align}
    &\delta w^{\mathsf{T}}\langle rr^{\mathsf{T}}\rangle_{\mathrm{te}}\delta w\notag\\
    &= b_{\mathrm{tr}}^{\mathsf{T}}b_{\mathrm{tr}} + b_{\mathrm{tr}}^{\mathsf{T}}a_{\mathrm{te}}b_{\mathrm{tr}} - 2b_{\mathrm{tr}}^{\mathsf{T}}a_{\mathrm{tr}}b_{\mathrm{tr}}\notag\\
    & \quad +O_p(T_\mathrm{tr}^{-2}) + O_p(T_{\mathrm{tr}}^{-3/2}T_{\mathrm{te}}^{-1/2}). 
\end{align}
Since the training and test datasets are independent and $\langle (a_{\mathrm{te}})_{ij}\rangle_u = 0$, we have
\begin{align}
    \langle b_{\mathrm{tr}}^{\mathsf{T}}a_{\mathrm{te}}b_{\mathrm{tr}}\rangle_u = \langle b_{\mathrm{tr}}^{\mathsf{T}}\langle a_{\mathrm{te}}\rangle_ub_{\mathrm{tr}}\rangle_u = 0. 
\end{align}
Using the same argument as in the covariance-based IPC analysis, $\langle b_{\mathrm{tr}}^{\mathsf{T}} a_{\mathrm{tr}}b_{\mathrm{tr}}\rangle_u = O(T_{\mathrm{tr}}^{-2})$ holds, and 
\begin{align}
    \langle C_{\mathrm{te}}(z\mid r)\rangle_u = C(z\mid r) - \frac{\Delta(z\mid r)}{T_{\mathrm{tr}}\langle z^2\rangle_u} + O(T^{-2}_{\mathrm{tr}}), 
\end{align}
where $\Delta(z\mid r) := \langle \epsilon^2\hat{r}^{\mathsf{T}}\hat{r}\rangle_u\ge 0$. 
Thus, the train/test split formulation of the IPC has a negative finite-sample bias at the order of $1 / T_{\mathrm{tr}}$. 

Since the objective of CROP is to estimate the underlying unperturbed IPC, we consider overestimation to be less desirable than underestimation. 
For this reason, we adopt the train/test formulation, which exhibits a negative finite-sample bias. 
Moreover, any negative estimates can simply be clipped to zero, avoiding the need for an additional threshold parameter. 
In addition, since the train/test IPC exhibits a finite-sample bias of the same order as the covariance-based IPC, the same bias-corrected estimator, $\hat{C}_T = 2C_T - C_{T/2}$ can be used to cancel the leading bias term. 
However, based on needs and priorities, a bias-corrected covariance based CROP may yield a reconstruction with lower NAE and AE, at the cost of some spurious IPCs.

\section{Finite-sample estimation of the noise-free information processing capacity\label{sec:traintest}}
In Sec.~\ref{subsec:proposed}, we propose the CROP method for estimating the noise-free IPC. 
The discussion there assumes that expectations over the random variables $u$ and $v$ can be evaluated exactly. 
In practice, however, these expectations must be approximated using finite-length time series. 
In this section, we describe a CROP method for finite data, where the time series is divided into training and test sets. 

Let $T_{\mathrm{tr}}$ and $T_{\mathrm{te}}$ denote the data length used for training and testing, respectively. 
We define empirical expectations over the training and test sets as $\langle\cdot\rangle_{\mathrm{tr}}$ and $\langle\cdot\rangle_{\mathrm{te}}$, respectively. 
In this train/test split setting, the optimal readout-weight is estimated using the training dataset, and the resulting weight is used to evaluate the IPC on the test dataset. 
Specifically, the IPC on the test set is given by
\begin{align}
    C_{\mathrm{te}}(z\mid r) = 1 - \frac{\langle (w_{\mathrm{tr}}^{*\mathsf{T}} r-z)^2\rangle_{\mathrm{te}}}{\langle z^2\rangle_{\mathrm{te}}}, 
\end{align}
where the optimal weight vector $w^*_{\mathrm{tr}}$ is obtained by minimizing the mean squared error (MSE) on the training set, 
\begin{align}
    w^*_{\mathrm{tr}} = (\langle rr^{\mathsf{T}}\rangle_{\mathrm{tr}})^{-1}\langle zr\rangle_{\mathrm{tr}}. 
\end{align}
Substituting this gives 
\begin{align}
    &C_{\mathrm{te}}(z\mid r)\\ &= 1 - \frac{\left\langle\left(\langle zr\rangle_{\mathrm{tr}}^{\mathsf{T}}(\langle rr^{\mathsf{T}}\rangle_{\mathrm{tr}})^{-1}r - z\right)^2\right\rangle_{\mathrm{te}}}{\langle z^2\rangle_{\mathrm{te}}}\notag\\
    &= \frac{2\langle zr\rangle_{\mathrm{tr}}^{\mathsf{T}}(\langle rr^{\mathsf{T}}\rangle_{\mathrm{tr}})^{-1}\langle zr\rangle_{\mathrm{te}}}{\langle z^2\rangle_{\mathrm{te}}} \notag\\
    & \quad- \frac{\langle zr\rangle_{\mathrm{tr}}^{\mathsf{T}}(\langle rr^{\mathsf{T}}\rangle_{\mathrm{tr}})^{-1}\langle rr^{\mathsf{T}}\rangle_{\mathrm{te}}(\langle rr^{\mathsf{T}}\rangle_{\mathrm{tr}})^{-1}\langle zr\rangle_{\mathrm{tr}} }{\langle z^2\rangle_{\mathrm{te}}}.
\end{align}
In the infinite-data limit, the statistical distinction between the training and test sets vanishes because the sample covariance matrices of both sets converge to the same true underlying distribution. 
Consequently, the above expression coincides with Eq.~\eqref{eq:czr}. 

Next, we consider the case where additive noise is present. 
Recall that 
\begin{align}
    \langle z\xi\rangle_{u,v} = 0
\end{align}
for any noise-independent function
\begin{align}
    z\in\mathcal{H}_u. 
\end{align}
Therefore, in the infinite-data limit and thanks to ergodicity, the cross terms between the target and the noise vanish exactly. 
With finite data, however, this orthogonality holds only approximately in empirical averages. 
Consequently, we have
\begin{align}
    \langle zr'\rangle_{\mathrm{tr}} &= \langle z(\bar{r} + \xi)\rangle_{\mathrm{tr}}\notag\\
    &\approx \langle z\bar{r}\rangle_{\mathrm{tr}}, \\
    \langle zr'\rangle_{\mathrm{te}} &= \langle z(\bar{r} + \xi)\rangle_{\mathrm{te}}\notag\\
    &\approx \langle z\bar{r}\rangle_{\mathrm{te}}, \\
    \langle r'r'^{\mathsf{T}}\rangle_{\mathrm{tr}} &= \langle (\bar{r} + \xi)(\bar{r} + \xi)^{\mathsf{T}}\rangle_{\mathrm{tr}}\notag\\
    &\approx \langle \bar{r}\bar{r}^{\mathsf{T}}\rangle_{\mathrm{tr}} + \langle \xi\xi^{\mathsf{T}}\rangle_{\mathrm{tr}}, \\
    \langle r'r'^{\mathsf{T}}\rangle_{\mathrm{te}} &= \langle (\bar{r} + \xi)(\bar{r} + \xi)^{\mathsf{T}}\rangle_{\mathrm{te}}\notag\\
    &\approx \langle \bar{r}\bar{r}^{\mathsf{T}}\rangle_{\mathrm{te}} + \langle \xi\xi^{\mathsf{T}}\rangle_{\mathrm{te}}
\end{align}

For reference, the unperturbed IPC to be estimated is given by
\begin{align}
    &C_{\mathrm{te}}(z\mid \bar{r}) \notag\\&= \frac{2\langle z\bar{r}\rangle_{\mathrm{tr}}^{\mathsf{T}}(\langle \bar{r}\bar{r}^{\mathsf{T}}\rangle_{\mathrm{tr}})^{-1}\langle z\bar{r}\rangle_{\mathrm{te}}}{\langle z^2\rangle_{\mathrm{te}}} \notag\\
    &\quad- \frac{\langle z\bar{r}\rangle_{\mathrm{tr}}^{\mathsf{T}}(\langle \bar{r}\bar{r}^{\mathsf{T}}\rangle_{\mathrm{tr}})^{-1}\langle \bar{r}\bar{r}^{\mathsf{T}}\rangle_{\mathrm{te}}(\langle \bar{r}\bar{r}^{\mathsf{T}}\rangle_{\mathrm{tr}})^{-1}\langle z\bar{r}\rangle_{\mathrm{tr}} }{\langle z^2\rangle_{\mathrm{te}}}.
\end{align}
The finite-data counterpart of the CROP method is defined as 
\begin{align}
    &\tilde{C}_{\mathrm{te}}(z) \notag\\
    &:= \frac{2\langle zr'\rangle_{\mathrm{tr}}^{\mathsf{T}}S_{\mathrm{tr}}^{-1}\langle zr'\rangle_{\mathrm{te}} - \langle zr'\rangle_{\mathrm{tr}}^{\mathsf{T}}S_{\mathrm{tr}}^{-1}S_{\mathrm{te}}S_{\mathrm{tr}}^{-1}\langle zr'\rangle_{\mathrm{tr}} }{\langle z^2\rangle_{\mathrm{te}}}, 
\end{align}
where 
\begin{align}
    S_{\mathrm{tr}} := \sum_{l=1}^L \langle y_lr'\rangle_{\mathrm{tr}}\langle y_lr'\rangle_{\mathrm{tr}}^{\mathsf{T}}, 
\end{align}
and
\begin{align}
    S_{\mathrm{te}} := \sum_{l=1}^L \langle y_lr'\rangle_{\mathrm{te}}\langle y_lr'\rangle_{\mathrm{te}}^{\mathsf{T}}. 
\end{align}
Applying the above approximations, we obtain
\begin{align}
    S_{\mathrm{tr}}&\approx \sum_{l=1}^L\langle y_l\bar{r}\rangle_{\mathrm{tr}}\langle y_l\bar{r}\rangle_{\mathrm{tr}}^{\mathsf{T}}\notag\\
    &\approx \langle \bar{r}\bar{r}^{\mathsf{T}}\rangle_{\mathrm{tr}},
\end{align}
and
\begin{align}
    S_{\mathrm{te}}&\approx \sum_{l=1}^L\langle y_l\bar{r}\rangle_{\mathrm{te}}\langle y_l\bar{r}\rangle_{\mathrm{te}}^{\mathsf{T}}\notag\\
    &\approx \langle \bar{r}\bar{r}^{\mathsf{T}}\rangle_{\mathrm{te}}. 
\end{align}
Substituting these approximation into the definition of $\tilde{C}_{\mathrm{te}}(z)$ yields 
\begin{align}
    &\tilde{C}_{\mathrm{te}}(z) \notag\\
    &\approx  \frac{2\langle zr'\rangle_{\mathrm{tr}}^{\mathsf{T}}S_{\mathrm{tr}}^{-1}\langle zr'\rangle_{\mathrm{te}} - \langle zr'\rangle_{\mathrm{tr}}^{\mathsf{T}}S_{\mathrm{tr}}^{-1}S_{\mathrm{te}}S_{\mathrm{tr}}^{-1}\langle zr'\rangle_{\mathrm{tr}} }{\langle z^2\rangle_{\mathrm{te}}} \notag\\
    &= C_{\mathrm{te}} (z\mid \bar{r}). 
\end{align}


\bibliographystyle{apsrev4-2}
\bibliography{apssamp}

\end{document}